\documentstyle[draft,psfig]{mn}
\newcommand{\mnref}[1]{\hangindent=0.5in \hangafter=1 #1 \par}

\newcommand{\mn}{MNRAS}

\newcommand{\aj}{AJ}
\newcommand{\apj}{ApJ}
\newcommand{\apjs}{ApJS}
\newcommand{\aaa}{A\&A}
\newcommand{\aas}{A\&AS}

\newcommand{\Lsolar}{\mbox{\,$\rm L_{\odot}$}}
\def\gs{\mathrel{\raise1.16pt\hbox{$>$}\kern-7.0pt
\lower3.06pt\hbox{{$\scriptstyle \sim$}}}}
\def\ls{\mathrel{\raise1.16pt\hbox{$<$}\kern-7.0pt
\lower3.06pt\hbox{{$\scriptstyle \sim$}}}}

\title{Spectropolarimetry of a Complete Infrared Selected Sample
of Seyfert 2 Galaxies}
\author[S.L. Lumsden, C.A. Heisler, J.A. Bailey, J.H. Hough \& S. Young]
{S.L. Lumsden$^{1,2}$, C.A. Heisler$^3$, J.A. Bailey$^2$, J.H. Hough$^4$
and S. Young$^4$ \\
{}$^1$ {\em Department of Physics and Astronomy,
University of Leeds, Leeds LS2 9JT, UK}\\
{Email -- sll@ast.leeds.ac.uk}\\
{}$^2$ {\em Anglo-Australian Observatory, PO Box 296, Epping, NSW 1710,
Australia}\\
{Email -- jab@aaoepp.aao.gov.au}\\
{}$^3$ {\em Mount Stromlo and Siding Spring Observatories, Private Bag,
Weston Creek P.O., Weston, ACT 2611, Australia\protect{\thanks{Charlene
Heisler died after a long struggle with ill-health on October 28 1999.
She will be greatly missed by all who knew her.}}}\\
{}$^4$ {\em Department of Physical Sciences, University of Hertfordshire,
Hatfield, Hertfordshire, AL10 9AB, UK}\\
{Email -- jhh@star.herts.ac.uk,
sy@star.herts.ac.uk}\\
}
\begin{document}

\label{firstpage}

\maketitle

\begin{abstract}
We report the results of a spectropolarimetric survey of a complete far
infrared selected sample of Seyfert 2 galaxies.  We have found polarized broad
H$\alpha$ emission in one new source, NGC5995.  In the sample as a whole, there
is a clear tendency for galaxies in which we have detected broad H$\alpha$ in
polarized light to have warm mid--far infrared colours (F$_{60\mu{\rm
m}}/$F$_{25\mu{\rm m}}\ls4$), in agreement with our previous results.  However,
a comparison of the optical, radio and hard x-ray properties of these systems
leads us to conclude that this is a secondary consequence of the true mechanism
governing our ability to see scattered light from the broad line region.  We
find a strong trend for galaxies showing such emission to lie above a critical
value of the relative luminosity of the active core to the host galaxy (as
measured from the [OIII] 5007\AA\ equivalent width) which varies as a function
of the obscuring column density as measured from hard x-ray observations.  The
warmth of the infrared colours is then largely due to a combination of the
luminosity of the active core, the obscuring column and the relative importance
of the host galaxy in powering the far infrared emission, and not solely
orientation as we inferred in our previous paper.  Our data may also provide an
explanation as to why the most highly polarized galaxies, which appear to have
tori that are largely edge-on, are also the most luminous and have the most
easily detectable scattered broad H$\alpha$.
\end{abstract}


\section{Introduction}

In the standard model for active galaxies (hereafter AGN), hard continuum
radiation from the accretion disk around a massive black hole ionises the
surrounding gas, which can be divided into two components: the broad line
region (BLR) and the narrow line region (NLR), at distances from the core of
$<1$pc and $<1$kpc, respectively (cf Osterbrock 1993).  For the case of
Seyfert galaxies, the classification into Seyfert 1 or 2 depends solely on the
presence or absence of considerably broadened (several thousand kms$^{-1}$)
permitted lines: ie, whether or not we can see the BLR in terms of the standard
model.  The simplest form of `unification' for Seyfert galaxies invokes a dusty
torus to obscure the light from the BLR in the Seyfert 2s, with the sole free
parameter being inclination (see, for example, Antonucci 1993 for a 
review of unification schemes for Seyfert galaxies).

The firmest evidence in favour of this simple unified model comes from optical
spectropolarimetry and x-ray spectroscopy.  Spectropolarimetry reveals
scattered broad permitted lines (eg.\ Antonucci and Miller 1985,
Miller and Goodrich 1990).  This can only come from a `hidden' BLR (hereafter
HBLR), that is obscured from our direct line of sight.  Similarly, x-ray data
for Seyfert 2s show evidence for either an absorbed Seyfert 1 like continuum in
the hard x-ray band, or when the extinction is sufficiently large, a scattered
spectrum characterised by an Fe K$\alpha$ line with large equivalent width
(see, for
example, the recent summary of the results from the ASCA satellite by Turner et
al.\ 1997a,b).  The evidence that at least some Seyfert 2s are in fact
`mis-aligned' Seyfert 1s is therefore compelling.  However, previous samples in
both the optical and x-ray have been selected in a fairly ad-hoc fashion.  For
the x-ray the limitation was simply sensitivity, so only the brightest x-ray
sources, which also tend to be the least obscured, were studied.  For the
optical spectropolarimetry, sources were often initially chosen on the basis of
their known high continuum polarization (eg.\ Miller and Goodrich 1990) or from
{\em ad hoc} samples drawn from the literature (eg.\ Tran, Miller and Kay 1992,
Kay 1994, Young et al.\ 1996a).  In addition, with the exception of Kay (1994),
who did not cover H$\alpha$ in the spectral range of her data somewhat limiting
its usefulness, and Young et al.\ (1996a), non-detections of HBLRs
were rarely published.

Therefore it is not possible to use the published results on optical
spectropolarimetry to test the unified model in a meaningful statistical sense,
or, indeed, to determine if the simplest form of unification is actually the
best match.  In order to interpret whether the detection of an HBLR relates to
orientation, or perhaps to other properties of the host galaxies, we must first
understand how the initial galaxy sample is chosen.  We therefore undertook a
new survey in which we obtained optical spectropolarimetry of a sample with
well understood far infrared properties.  As discussed in Section 2, the far
infrared luminosity is likely to be a relatively extinction (and hence
orientation) independent measure of the combined AGN and host galaxy
luminosity.  In addition, there was a notable trend shown in early detections
of HBLRs for those galaxies to have `warm' mid--far infrared colours (eg Inglis
et al.\ 1993).  We therefore wished to determine if this was still true for a
larger sample, and that the AGN with cooler colours did {\em not} show evidence
for HBLRs.  The initial results of that survey were reported in Heisler,
Lumsden and Bailey (1997: hereafter Paper 1).  There we reported an apparent
trend for HBLRs to only appear in those AGN where F$_{60\mu{\rm
m}}/$F$_{25\mu{\rm m}}<4$.  Since the mid-far infrared emission in Seyfert
galaxies is dominated by thermal emission from dust, this trend is consistent
with the unified model, since low values of the F$_{60\mu{\rm
m}}/$F$_{25\mu{\rm m}}$ ratio imply warmer dust temperatures.  Therefore we
concluded that galaxies with warm mid-far infrared colours had tori that were
closer to face-on, and hence more likely to show scattered radiation from the
BLR.

However, our initial survey did suffer from some problems relating to the
actual sample observed (see Section 2 for details), and was simply too small to
test for other dependencies in the data that might mimic the effect that
orientation would have.  In particular, it is possible that star formation may
play an increasingly important role in the AGN with cooler mid--far infrared
colours, since the dust emission surrounding HII regions typically peaks nearer
100$\mu$m.  Alexander (2001) suggested that the hard x-ray emission from the
sample given in Paper 1 is consistent with such a picture, since the AGN with
both warm and cool colours show a similar spread in implied neutral hydrogen
column density.  He inferred from this that the mid-far infrared colours
related more to the presence or absence of significant star formation rather
than orientation.  In addition he noted, as had Tran et al.\ (1999),
that there was an apparent trend for the HBLRs to have higher [OIII]
5007\AA/H$\beta$ ratios, and indeed for this ratio to be correlated with
mid--far infrared colour.  Again, this may be consistent with enhanced
star formation in the galaxies with cooler colours, since the starburst
will enhance H$\beta$ but have relatively little effect on [OIII] emission
in most cases.

This paper therefore presents the results of an extended survey of far infrared
selected AGN.  Since Paper 1 we have also refined our selection to exclude
objects whose actual classification is in doubt (see Section 2.5).  We will
present the full spectropolarimetric data, including non-detections, for all
objects for which we acquired data in a later paper.  We will also present
detailed modelling of all aspects of the observed polarimetry rather than just
a simple consideration of the presence or absence of scattered broadened
H$\alpha$ emission in that paper.

\section{The Sample and Observing Details}
\subsection{Selection Method}
Any test of AGN unification must use a sample selected on a known isotropic
property since the unified model implies an orientation dependence for the type
of Seyfert seen.  It is important to avoid the error, often seen in the
literature, that all Seyfert 2's are identical.  If the unified model is
correct, Seyfert 2s must have a wide range of orientation of the AGN axis with
respect to our line of sight, from narrowly mis-aligned so that the Seyfert 1
core is only just out of view, to those where the broad and narrow line regions
are almost entirely in the plane of the sky.  It is highly unlikely that the
observed properties of Seyfert 2's seen at widely different orientation angles
are the same.  Therefore it is not sufficient to pick a sample of Seyfert 2
galaxies at random from the literature, as any such sample will almost
certainly be biased with respect to orientation.

In practice, a combination of extinction, opacity effects, and varying
beam-size at different wavelengths, makes it impractical to select on any
property related solely to the active nucleus.  It is possible though to select
based on intrinsic properties of the galaxy as a whole, in a wavelength regime
where the active nucleus itself is likely to be important.  Examples of this
type include far infrared (hereafter FIR), cm radio, and global optical
emission.  Each of these has its own particular advantages and disadvantages,
but we initially chose a FIR selection since there is a complete FIR selected
galaxy database readily available.  

The nature of the FIR emission in Seyfert galaxies was the subject of
considerable speculation until relatively recently.  The turnover in the
spectral energy distribution beyond 100$\mu$m was seen as evidence for a
thermal origin (Chini, Kreysa and Biermann 1989).  More recent ISO photometry
has helped to confirm that this picture is correct for most if not all nearby
radio quiet AGN (eg Perez Garcia and Rodriguez Espinosa 2000).  There have been
many attempts to model the properties of the dust emission around the active
nucleus itself in the context of the unified model (eg Pier and Krolik 1992,
Laor and Draine 1993, Granato and Danese 1994, Efstathiou and Rowan-Robinson
1995).  These models suggest that the 60--100$\mu$m emission from any AGN is
largely unaffected by orientation.

Some fraction of the emission at these longer wavelengths could be due to the
host galaxy rather than the core however.  Star formation in particular gives
rise to significant FIR emission.  It is possible however to distinguish the
main source of the FIR emission using simple colour-colour diagrams of the
observed IRAS fluxes (eg.\ Dopita et al 1998).  Figure 1 shows IRAS
colour-colour plots for the sample of galaxies considered by Kewley et al.\
(2000).  That sample shares similar FIR selection criteria to those used in
Paper 1 and this paper, but included all types of galaxy.  The data shown here
differ from that in Kewley et al.\ since we have revised the IRAS fluxes to use
the data from the IRAS Bright Galaxy Survey (BGS: Soifer et al. 1989 and
Sanders et al.\ 1995).  The solid line in the figures represents the track for
an increasingly reddened Seyfert galaxy, whereas the dashed line shows the
locus of known starburst galaxies (see Dopita et al.\ 1998).  Galaxies of mixed
excitation should lie between these lines (the dot-dashed line shows the
extreme mixing line between starburst and AGN activity).  Seyfert galaxies of
both types are shown in the left hand panel, and starbursts and LINERs in the
right, in order to show the clear differences in their observed properties.

\subsection{The Infrared Catalogue}
We used the catalogue of IRAS galaxies and the redshift survey of Strauss et
al.\ (1990, 1992) to identify our AGN sample.  We did not rely on the Rush et
al.\ (1993) catalogue used in Paper 1 for several reasons.  First, it suffers
badly from classification errors.  This meant we had to independently classify
all galaxies that met our selection criteria regardless of which source we used
for the IRAS data.  Secondly, the Strauss et al.\ catalogue is larger, and
complete to fainter flux limits, making it better for identifying a larger
parent sample of AGN.  Lastly, doubts have been raised about the accuracy of
the IRAS fluxes in the Rush et al.\ catalogue (Alexander \& Aussel 2000).

Strauss et al.\ (1992) carried out a spectroscopic survey of sources selected
from the IRAS Point Source Catalogue (PSC v2.0) by Strauss et al.\ (1990), to
determine redshifts of the galaxies, and remove non-extragalactic sources.  The
actual IRAS fluxes they quote are not drawn solely from the PSC however.  Where
the PSC flags a source as extended or variable, Strauss et al.\ (1990) derived
ADDSCAN fluxes.  These should in principle be better than the PSC fluxes since
(i) they correctly sum all the flux from an extended object and (ii) they coadd
more of the raw IRAS data and are therefore of higher signal-to-noise.  Strauss
et al.\ (1990) discuss this issue at greater length.  The final catalogue of
Strauss et al.\ (1992) contains most galaxies in the original PSC
down to a 60$\mu$m flux limit of 1.9Jy.

The IRAS fluxes actually quoted in this paper however are derived from a
variety of sources, since we sought the best possible data once the initial
selection was carried out.  We compared the fluxes in Strauss et al., the IRAS
Faint Source Catalogue Version 2 (FSC: Moshir et al.\ 1991), and the BGS
(Soifer et al.\ 1989, Sanders et al.\ 1995).  These all agree within the
quoted errors for point sources.  The BGS data are also ADDSCAN fluxes of the
original raw data, but with a higher degree of rejection of `bad' scans.  They
should therefore be more accurate than the Strauss et al.\ data.  Therefore,
for galaxies with F$_{60\mu{\rm m}}>5$Jy we used the fluxes from the BGS.  We
used the data of Strauss et al.\ (1990) for galaxies with F$_{60\mu{\rm
m}}<5$Jy where those data were derived from ADDSCAN fluxes.  Finally, for all
other galaxies we used the data quoted in the FSC (which again is produced from
coadding more of the raw IRAS data than the PSC).  The BGS data quote errors on
their fluxes, as does the FSC.  We assume the errors on the Strauss et al.\
ADDSCAN fluxes must be bound by the FSC errors.  

Our initial sample is based on those galaxies with good detections in the PSC
(ie quality flag set to 3) at 60$\mu$m.  They also have moderate or good
detections at 25 and 100$\mu$m (quality flag set to either 2 or 3).  We did not
set any restriction on the 12$\mu$m data.  However, because we do not use the
actual PSC data in our analysis, all objects in our final sample actually have
firm detections at all four IRAS wavebands.

\subsection{Detailed Selection Parameters}

The detailed selection criteria of our sample are as follows (some of these are
actually imposed by using the Strauss et al.\ catalogue):
$L_{FIR}>10^{10}$\Lsolar (where F$_{FIR}$ has its usual definition --
F$_{FIR}=1.26\left(2.58{\rm F}_{60\mu{\rm m}} + {\rm F}_{100\mu{\rm
m}}\right)\times10^{-14}$Wm$^{-2}$); F$_{60\mu{\rm m}}/$F$_{25\mu{\rm m}}<8.5$
which matches most known Seyferts (de Grijp et al.\ 1992) without requiring us
to classify the large population of normal galaxies with cooler IRAS colours;
F$_{60\mu{\rm m}}>3$Jy; Galactic latitude $|b|>25^\circ$; declination less than
+70$^\circ$; and classification as a Seyfert 2 (from our own spectroscopy if
possible: see Section 2.5).  We note that we class Seyfert 1.8 and 1.9 galaxies
as Seyfert 2's for the purpose of this paper.  We use a value of
$H_0=75$kms$^{-1}$Mpc$^{-1}$ throughout.

These criteria are looser than those adopted in Paper 1.  Despite this, some of
the galaxies discussed there do not fall into the current sample.  These are
NGC34 (more accurately classified as a LINER), NGC7496 and NGC7590 (both
L$_{\rm FIR}<10^{10}$\Lsolar).  None of these objects showed evidence for an
HBLR.

Table 1 lists the final sample of 28 galaxies that satisfy all our criteria.
We do not have spectropolarimetry for 4 of these objects.  All have cool
colours, with F$_{60\mu{\rm m}}/$F$_{25\mu{\rm m}}>6$.  A detailed discussion
of these sources is deferred until the Appendix, though we will note, where
appropriate, the effect they have on our analysis.

\subsection{Observations}

Full details of the observing procedures and data reduction will be given in a
future paper.  With the exception of those objects for which we have used data
available in the literature, all spectropolarimetry was obtained at either the
AAT or WHT, on the nights of 16, 17 and 18 August 1995 (AAT), 29 July 1996
(WHT), 26, 27 and 28 August 1997 (AAT) and 7, 8, 9 and 10 March 1998 (WHT).
Only the data from 1998 were obtained with the slit at the parallactic angle
(mainly for reasons of practical convenience when carrying out
spectropolarimetry).  We centred the slit on the brightest part of the galaxy
at approximately the V band.  The conditions were photometric for the all
except the night of 18 August 1995.  Total exposure times were approximately
two hours for all sources observed at the AAT, and one hour for the sources
observed with the more efficient ISIS spectrograph at the WHT.  Ideally, the
observations would seek to achieve a level consistent with that expected of any
broad line in the scattered light (cf Section 3.1).  In practice, instrumental
and other effects limited us to a fixed small error in the polarization of
$\sim0.3$\%.  The fainter sources, and some that were observed in adverse
weather conditions, have larger observed errors.  Fuller details of the 
errors will be given in the paper presenting the full data.

For the AAT data, we used the RGO Spectrograph together with a Tektronix
1024$^2$ CCD.  For the WHT data, we used the red arm of the double beam
spectrograph ISIS with a Tektronix 1024$^2$ CCD.  A calcite prism was used in
both instruments to split the incoming beam into $e$ and $o$-rays together with
an aperture mask made up of discrete slit-lets.  A half-waveplate modulated the
incoming phase.  Four steps of the waveplate at 0$^\circ$, 45$^\circ$,
22.5$^\circ$ and 67.5$^\circ$ were required to derive the full set of Stokes
parameters for linearly polarized light.  The slit width was typically 2
arcseconds at the AAT, and 1--1.5 arcseconds at the WHT.  We extracted data
from a region of between 3 and 4 arcseconds along the slit in most cases.  Sky
subtraction was achieved by nodding the object into an adjacent slit-let.  The
data were reduced in a standard fashion.  The effective spectral resolution is
between 7 and 9\AA\ for all data depending on the exact slit width used.  The
sky and object spectra for the four separate waveplate positions were extracted
then combined to give the final Q, U and I Stokes parameters.

\subsection{Spectral Classification}

We measured the relative line fluxes from our spectra by fitting Gaussian
profiles.  Where the actual observed profile is highly non-Gaussian,
we used a multiple component Gaussian fit to measure the total
flux, since we are not interested in the detailed line profile information
itself.  We also allowed for the possibility of stellar H$\beta$ absorption in
our spectra.  It is possible to overestimate the internal extinction, and
mis-classify starburst galaxies as AGN without this correction.  We felt it was
better to actually fit the absorption component directly rather than making an
average correction as has often been done in the past.  This will make little
difference for galaxies with an old underlying stellar population, where the
H$\beta$ absorption is weak in any event, but has a significant effect on those
galaxies which have a stellar continuum similar to that of an A star
(essentially those galaxies with a post-starburst population).


The extinction was derived using the Whitford reddening curve as parameterized
by Miller \& Mathews (1972), assuming the intrinsic H$\alpha$/H$\beta$ ratio is
3.1.  The extinction corrected line ratios for our own spectra, and for those
galaxies taken from the literature, are given in Table 2.  We note that the
derived $E(B-V)$ values may not be completely accurate since we did not observe
at the parallactic angle for most of our observations.  This is especially a
problem for strongly nucleated sources observed well away from the parallactic
angle, since a point source at the wavelength of H$\beta$ will appear to shift
by up to 1 arcsecond from its position at H$\alpha$ due to differential
atmospheric dispersion.  Similarly, such effects can lead to errors in line
fluxes that have been corrected for extinction.  By contrast, line ratios, as
long as they arise from the same component, are largely unaffected, since the
extinction correction essentially forces the data to have the `correct'
spectral form.  
The same arguments can also be made regarding data taken under
non-photometric conditions.   These caveats also apply to much of the data
presented in the literature, and should be borne in mind as a possible source
of error in the extinction corrected [OIII] 5007\AA\ line luminosities
quoted in Table 2 and used throughout this paper.

We classified each galaxy based on the diagnostic diagrams of Veilleux and
Osterbrock (1987).  We also considered whether the galaxy showed evidence of
HeII 4686\AA\ emission.  The high signal to noise of our spectropolarimetry
data means we are able to detect this line where it is present.  This line is,
by definition, not seen in LINER galaxies.  It is seen in some starburst
galaxies, where it arises in the winds of Wolf-Rayet stars (see, eg, Vacca and
Conti 1992).  However, the feature in Wolf-Rayet galaxies is much broader than
the other observed lines, whereas in all the galaxies we observed the width of
the 4686\AA\ line is consistent with the width of [OIII] 5007\AA, indicating it
also arises in the NLR.  

Four of the galaxies included in Table 2 do not have standard Veilleux \&
Osterbrock Seyfert classification.  We have kept these galaxies in our sample
on the basis of other reported characteristics.  0031--2142 is classed as a
Seyfert 1.8 by Moran et al.\ (1996) on the basis of weak broad H$\alpha$
emission.  Mrk~334 is classed as a Seyfert on the basis of a weak [FeX] line,
which cannot arise in a LINER.  0019--7926 shows weak HeII emission.  It also
seems possible that we have not observed the actual nucleus.  Both de Grijp et
al.\ (1992) and Vader et al.\ (1993) present spectra of this galaxy which have
a standard Seyfert classification on the basis of all the Veilleux \&
Osterbrock (1987) indices, as well as significantly stronger [OIII] 5007\AA\
emission than is present in our spectrum.  However, Kewley et al.\ (2001)
present data that are in agreement with ours.  Given the uncertainty we have
decided to include this object.  NGC~7582 shows weak HeII emission, and has
been observed to show transient broad H$\alpha$ (Arextaga et al.\ 1999).  We
tested for any effect that the inclusion of these galaxies had on the
statistical tests that we describe in Section 3.  There was no significant
difference in any case.  The only difference that their exclusion makes is to
the overall detection rate reported in Section 3.1, and we discuss that factor
there.  We also discuss, where appropriate, any other effect that the inclusion
of these galaxies may have.

\section{Results}
\subsection{The Polarimetry Data}
Only one of the galaxies newly reported in this paper, NGC5995, contains an
obvious HBLR.  Broad H$\alpha$ and H$\beta$ are clearly evident in the
polarized flux from this object (see Figure 2).  Here we have plotted the
Stokes flux from our observations (derived from the rotated Stokes parameter --
these quantities are a less biased estimate of the actual polarization and
polarized flux present since they are not positive biased -- see Miller,
Robinson and Goodrich 1988).  The measured full width at half maximum of the
broad H$\alpha$ component is approximately 2700kms$^{-1}$.  Although the data
has lower signal-to-noise near H$\beta$ we can also measure the
H$\alpha$/H$\beta$ ratio in our polarized flux spectrum.  We obtain a ratio of
18$\pm$6, which corresponds to $E(B-V)=1.7\pm0.4$ assuming an intrinsic ratio
of 3.1.  We can compare this with the extinction value given in Table 2, which
shows $E(B-V)=1.78$.  Given the similarity of these values it would appear that
most of the extinction lies between us and the scattering particles or narrow
line region, rather than between the AGN core and the scatterers.  Finally, one
other notable feature of the spectrum of NGC5995 is that it shows clear
evidence for strong stellar absorption lines.  This is rather unusual for HBLRs
as noted by Kay and Moran (1998).

Two other galaxies show weak features that may be the signature of an HBLR.
NGC5135 has a weak broad feature in the polarized flux, but no signature in the
polarization itself.  In addition, the residual broad feature left after
subtracting out the narrow line component is considerably redshifted with
respect to the systemic velocity (by more than 1000kms$^{-1}$).  This could be
due to the weakness of the feature however, since it is possible that we have
over-corrected for the narrow H$\alpha$ and [NII] components in the polarized
flux spectra, and hence removed the blue `wing' of the broad feature.  NGC5929
shows a weak broad feature in the polarization at H$\alpha$ but no evidence for
a broad component in the actual polarized flux.  Again, it is possible that a
weak broad line could be masked by the much stronger polarized narrow lines.
We do not therefore consider these features as detections of HBLRs in either
galaxy, but clearly both sources are worthy of further study to confirm whether
this is true.

We also considered whether we would expect to see an HBLR given the
signal-to-noise achieved in these observations.  We derived predicted broad
H$\alpha$ fluxes from the observed [OIII] 5007\AA\ fluxes, with due correction
for the difference in extinction between the two lines, but assuming that both
the [OIII] emission from the NLR and the scattered light from the BLR suffer
the same extinction along our line of sight.  We used two methods for deriving
the predicted flux.  We fitted the observed relation between [OIII] 5007\AA\
and broad H$\alpha$ for the Seyfert 1s observed by Stirpe (1990), and then
assumed that an average of 2\% of the BLR light is scattered into our direction
in Seyfert 2s (see Lumsden \& Alexander in preparation) for the first
prediction.  For the second we fitted the observed relation between the
scattered broad H$\alpha$ flux and the direct [OIII] 5007\AA\ flux in the HBLRs
studied by Tran (1995a), Young et al.\ (1996a) and ourselves.  The two results
agree to within a factor of 5.  We then derived an upper limit to the observed
scattered broad H$\alpha$ in our data from the Stokes flux, assuming the full
width at half maximum of the line was 3600kms$^{-1}$ (the average for the
Seyfert 1s in Stirpe 1990).  Where our limit is 3 times less than the predicted
value in both cases we assume that the non-detection is secure.  Where this
condition is satisfied by only one of the predicted fluxes, we consider the
non-detection is likely but flag the possibility that this is wrong in Table 2.
In two cases, NGC7479 and 1408+1347, our limit is above the predicted values,
and we cannot therefore rule out the presence of an HBLR.  These results are
broadly in line with the analysis of Alexander (2001) of the sample in Paper 1.
Further observations of all the galaxies for which we cannot absolutely rule
out the presence of an HBLR are clearly desirable.  For the rest of this paper,
however, we will assume that all of the galaxies in which we did not detect an
HBLR are in fact non-HBLRs.

In total 8 out of the sample of 24 observed galaxies show evidence for a HBLR.
Assuming the 4 unobserved galaxies with cooler IRAS colours are not HBLRs, this
gives an overall detection rate of 29\%.  If we exclude the 4 galaxies with
predominantly intermediate or LINER types discussed in section 2.5, the
detection rate would be 8 out of 21 observed, and a total sample of 24.  This
gives a rate of 33\%.  Both of these are close to the value reported by Moran
et al.\ (2000) based on an initial volume limited optical selection of 35\%.
The similarity probably indicates our initial selection is largely unbiased and
provides a fair representation of the full range of nearby Seyfert 2s.

\subsection{Optical and Infrared Diagnostics}
Figure 3 shows where the HBLRs in our sample fall as a function of IRAS colour.
Our expanded survey indicates that the trends we reported in Paper 1 still
hold.  There is a clear correlation between the mid-far infrared flux ratio and
the ability to detect an HBLR.  Galaxies with warmer colours (F$_{60\mu{\rm
m}}/$F$_{25\mu{\rm m}}\ls4$) almost uniformly show evidence for an HBLR: the
two exceptions are the optically faint galaxies 0019--7926 and 0425--0440.  As
noted in Section 2.5 we may not have observed the nucleus of 0019--7926.  If we
have indeed misclassified this particular galaxy the net effect on our results
is small as discussed in more detail in the Appendix.  0425--0440 is one of the
galaxies in which we cannot rule out the possibility of a misclassification 
due to insufficient signal-to-noise.

Figure 3 can also be compared directly with Figure 1.  It is clear that there
is a greater spread in the Seyfert sample under consideration here than in the
Kewley et al.\ (2000) sample.  This is largely due to the inclusion of the
galaxies which do not have standard Seyfert classifications (see Section 2.5),
the low luminosity Seyferts in NGC5194 (M51) and NGC7172, and the inclusion of
the galaxies for which we do not have spectropolarimetry (for which we are
generally reliant on only one literature source for classification).  In most
of these cases it is likely from the IRAS data alone that the active core is
not the dominant source of luminosity, since they follow the same track as the
starburst and LINER population in Figure 1.  By contrast the HBLRs uniformly
follow the reddened Seyfert 1 track.  The other striking aspect of the HBLRs
from Table 2 is that they are always classified as Seyferts on the three
standard Veilleux \& Osterbrock classification diagrams.  The HBLRs and the
starburst/LINER dominated galaxies represent the two extremes in our data.  The
rest of the non-HBLR population lies between the two, with some on the Seyfert
1 reddening track, and some on the starburst/LINER track.

We considered whether luminosity is a factor in the detection of the HBLRs.  A
Mann-Whitney U test shows that the distributions of L$_{FIR}$ for the HBLRs and
non-HBLRs are consistent at the 95\% level. Therefore L$_{FIR}$ is not a factor
in determining the presence of a HBLR, in agreement with our findings in Paper
1.  A different result is obtained if we look at the [OIII] 5007\AA\ line
luminosity, L$_{\lambda5007}$.  It is generally held that the luminosity in this line
is a good measure of the luminosity of the active core (Mulchaey et al.\ 1994;
Alonso-Herrero, Ward and Kotilainen 1997).  Our data suggest marginal support
for there being a difference between the HBLRs and non-HBLRs at the 10\%
significance level, in the sense that the HBLRs are more luminous on average.

If we look at how the IRAS colours behave as a function of L$_{\lambda5007}$, we find
that all galaxies with L$_{\lambda5007} >10^{8}$\Lsolar\ lie near the Seyfert 1
reddening line in Figure 3, regardless of whether they are HBLRs.  This last
point is confirmed by the weak correlation between L$_{\lambda5007}$ and IRAS colour
at a 5\% significance level shown in Figure 4(a): the more luminous sources
have warmer colours.  There is no such correlation between L$_{FIR}$ and colour
as shown by Figure 4(b).  

Another way to demonstrate the dependence of the IRAS fluxes on the AGN
luminosity is to consider what fraction of the mid infrared flux actually
arises near the active core.  We have used the small aperture 10$\mu$m data
presented in Giuricin, Mardirossian and Mezzetti (1995) and Maiolino et al.\
(1995) to determine how `compact' the IRAS 12$\mu$m emission is.  This is done
in the standard fashion by defining a compactness parameter, C, which is the
ratio of the small aperture 10$\mu$m flux and the IRAS 12$\mu$m flux scaled by
an appropriate colour correction (Devereux 1987).  The results are given in
Table 3.  Values of C near 1 indicate that the IRAS emission arises from within
the aperture of the 10$\mu$m beam.  Although only half of our sources have
small aperture 10$\mu$m data, the results are instructive.  All of the galaxies
with L$_{\lambda5007} >10^{8.5}$\Lsolar\ have C$>0.65$. Only NGC5135 has a value of
L$_{\lambda5007}<10^{8.5}$\Lsolar\ and a value of C larger than 0.65.  These results
are consistent with our findings above, since they indicate that
the galaxies with lower AGN luminosities are more likely to show significant
host galaxy emission in the mid infrared, which must also be true in the
FIR as well.  

We also used the equivalent width of the [OIII] line, W$_{\lambda5007}$, as a
measure of the ratio of the luminosities in the active core and the host
galaxy.  Figure 5 shows that W$_{\lambda5007}$ and L$_{\lambda5007}$ are correlated as
might be expected, but only at the 10\% significance level, and there is a
large scatter.  Galaxies with high values of W$_{\lambda5007}$ tend to be more
luminous, and to be more clearly classified as Seyferts (Table 2).  In this
sense, W$_{\lambda5007}$ is a good indicator of whether or not the AGN
dominates the optical spectrum and the overall spectral energy distribution.

In Figure 6(a) we show the relation between W$_{\lambda5007}$ and F$_{60\mu{\rm
m}}/$F$_{25\mu{\rm m}}$.  Those galaxies which have large equivalent widths
also show HBLRs.  Formally, the galaxies with and without HBLRs have
distributions of W$_{\lambda5007}$ that are different at the 99\% confidence
level.  There
are counter examples, since NGC5995 and 0518--2524 both have values of
W$_{\lambda5007}$ which are more typical of the bulk of our sample.  
Those galaxies with the largest W$_{\lambda5007}$ also have the warmest
colours.   Again this is consistent with the picture outlined above 
with regard to AGN luminosity and mid--far infrared colour.

As noted by Tran et al.\ (1999) and Alexander (2001) there is a clear trend for
Seyfert 2s with HBLRs to have both warm IRAS colours and evidence for high
excitation through the [OIII] 5007\AA\ to H$\beta$ ratio.  Again, this is true
for most of our sample.  Figure 6(b) shows the extinction corrected
[OIII] 5007\AA\ to
H$\beta$ ratio as a function of F$_{60\mu{\rm m}}/$F$_{25\mu{\rm m}}$.
Clearly, most of the HBLRs tend to have higher apparent excitation than the
sample as a whole, and the two subsets are indeed statistically different at
the 99.9\% confidence
level.  This is not due to a dependence of the excitation on AGN
luminosity as there is no correlation between the [OIII] 5007\AA\ to H$\beta$
ratio and L$_{\lambda5007}$, nor is there a tendency for the more luminous AGN to have
larger HeII 4686\AA\ to [OIII] 5007\AA\ line ratios as might be expected if
there was such a correlation.  A large [OIII] 5007\AA\ to H$\beta$ ratio is not
a necessary requirement for an HBLR however: NGC5995 has a ratio $\sim6$,
closer to that of the non-HBLRs with standard Seyfert classifications in Table
2 (cf NGC5135 and NGC7130).  Nor is a large [OIII] 5007\AA\ to H$\beta$ ratio a
clear sign of an HBLR, or of warm IRAS colours, since there are galaxies in our
sample with cool colours and large [OIII] 5007\AA\ to H$\beta$ ratio and
without an HBLR.

\subsection{Radio and X-Ray Properties of the Sample}
The two other wavelengths with greatest power for examining the properties of
the active core are cm radio and hard (2--10keV) x-ray.  Shorter wavelength
radio emission, and lower energy x-ray emission, can be due to thermal
emission from star formation if present.  We therefore searched the literature
for both integrated and core radio fluxes at 2.3GHz, as well as the x-ray
fluxes in the 2--10 keV band, and the neutral hydrogen column densities
inferred from these.  The data we used in our study are summarised in Table 3.
The presence of many upper limits in the following requires the use of the
techniques of survival analysis in comparing samples.  We have made use of the
software package ASURV, version 1.1 (La Valley, Isobe \& Feigelson 1992), which
implements methods for univariate and bivariate problems, as outlined in
Feigelson \& Nelson (1985) and Isobe, Feigelson \& Nelson (1986).

One caveat should be made with regard to the core radio fluxes.  Most of these
values were obtained using the Parkes Tidbinbilla Interferometer (PTI), which
linked the 64m Parkes antenna with the 70m Tidbinbilla antenna over a 275~km
baseline.  Most were obtained in snapshot mode, so if there is asymmetric
structure on the scales to which the PTI is sensitive ($\sim0.1$~arcseconds)
the flux will be underestimated if the PTI beam is not aligned with that
structure.  A comparison of the galaxies observed by both Sadler et al.\ (1995)
and Thean et al.\ (2000) at 8.4GHz VLA show this effect clearly, even for some
systems which are only marginally resolved by the PTI.  However, most of our
sources are likely to have unresolved cores, rather than more complex radio jet
structures, and the PTI observations cover most of our sample, unlike the
available VLA observations.  Thean et al.\ (2000) did report detections of some
compact cores at 8.4GHz for sources that have only upper limits in the PTI
observations.  We have not included these since it is clear from work such as
Sadler et al.\ (1995) that the spectral index of the core varies widely,
whereas the general emission from the galaxy is reasonably well matched to an
index near $-0.75$.  Furthermore, the 8.4GHz emission can be dominated by star
formation.  We note here though that at 8.4GHz the core in 0518--2524
contributes more than 50\% of the total flux, whereas the cores in NGC4388 and
NGC5194 contribute less than 10\%.  For comparison the 2.3GHz limits on all 3
are less than 10\%.

There is no significant difference in the inferred core or total radio power
between the galaxies showing evidence for an HBLR and those that do not.
However, there is a weak, though not statistically significant, trend for the
HBLRs to have slightly higher core radio power.  Unfortunately the limited
amount of data available on the core radio fluxes and the large number of upper
limits make it difficult to say whether this is real.  There is also weak
evidence (significance only 15\% however) for the galaxies containing HBLRs to
have higher ratios of core radio flux to far infrared flux (Figure 7a), but not
higher ratios of integrated radio flux to far infrared flux (Figure 7b).  These
plots are essentially measures of how important star formation is to the radio
and far infrared emission (see Helou, Soifer \& Rowan-Robinson 1985).
Galaxies in which the ratio of the total radio to FIR flux is high are likely
to be AGN dominated.  Figure 7 tends to indicate that most, but not all, HBLRs
lie in systems in which the AGN dominates.

We also examined how well the radio luminosities correlated with the optical
and infrared data.  L$_{\lambda5007}$ correlates very well with both the core and
total radio power at better than the 99\% confidence level.  By contrast,
L$_{FIR}$ correlates with the total radio power at the 99\% confidence level,
but with the core radio power at only the 95\% confidence level.  The FIR-radio
correlation for all radio quiet galaxies is well known (eg de Jong et al.\
1985).  The fact that Seyferts show a larger scatter in this relationship when
compared to starbursts is also well known (Norris, Allen \& Roche 1988, Sopp \&
Alexander 1991).  A discussion of the relative contribution of the AGN core to
the total flux can be found in Roy et al.\ (1994, 1998) and Heisler et al.\
(1998), who note that the scatter seen in the FIR--radio correlation is largely
due to those galaxies with powerful radio cores.  Our results are consistent
with these findings.  In particular, the tight correlation between L$_{\lambda5007}$
and the core radio power strengthens the case for L$_{\lambda5007}$ being a good
measure of the AGN luminosity, and the poorer correlation between L$_{FIR}$ and
the core radio power indicates that L$_{FIR}$ is not solely a measure of the
AGN luminosity.

The 2--10keV x-ray band is important since it allows a clean measure of the
intrinsic AGN flux, and many nearby AGN have been observed spectroscopically 
in that band by ASCA.  For most of the galaxies in our sample the active core
will be completely obscured below 2keV.  The x-ray data can be used to derive
the neutral hydrogen column densities to the central source.  For sources with
$N_H<10^{24}$cm$^{-2}$, the values are derived from the observed photoelectric
cut-off.  Sources with $N_H>10^{24}$cm$^{-2}$, often described as Compton
thick, are obscured at 10keV, but may show reflection dominated spectra.  The
`observed' extinction in these sources is therefore low since none of the
direct flux actually reaches us.  The inferred high column density is
determined from other properties, such as the ratio of the observed
2--10keV flux with extinction corrected [OIII] 5007\AA\ flux, which increases
as the column density decreases, and the 6.4keV Fe K$\alpha$ line equivalent
width, which increases as the scattered component becomes more dominant.  In a
few cases, Compton thick sources have been observed with the BeppoSAX
satellite, which can detect emission in the 10-100keV band as well.  This
allows more stringent limits to be placed in these cases (eg.\ Matt et al.\
2000).  The actual values as taken from the literature are given in Table 3,
along with the original references.

For two of the objects in Table 3 we have used our own reductions of archived
ASCA data.  We only used the GIS2 and GIS3 data, and combined and grouped these
data so that after rebinning there were at least 20 counts in each bin.  We
fitted absorbed power-laws plus a Gaussian (to represent the FeK$\alpha$ line)
to the data in the 2--10keV range, analogous to the procedure in Bassani et
al.\ (1999).  This is actually a good fit to the observed spectrum of NGC5995,
which is relatively bright and well detected by ASCA.  The equivalent width of
the best fit line is 240$^{+240}_{-160}$eV, and the 
photon index, $\Gamma=1.8^{+0.14}_{-0.11}$.  The rather low obscuring column
seen is therefore intrinsic, and not due to us only seeing the source through
scattered radiation.  The ASCA data for IC3639 are of relatively low
signal-to-noise, and allowed only a crude estimate of the equivalent width, of
4200$^{+8000}_{-4200}$eV, if the photon index was fixed at 2, and the line
centre fixed at 6.34keV.  However, this is consistent with the object being
Compton thick, as also found by Risaliti, Maiolino and Salvati (1999).  We use
their value of N$_H$, derived from BeppoSAX data, since they determine a larger
lower limit.  

In Figure 8(a) we plot W$_{\lambda5007}$ against $N_H$.  There is a clear
separation between the HBLRs and the non-HBLRs, in the sense that only the
galaxies with the largest value of W$_{\lambda5007}$ at any given obscuration
show HBLRs.  Figure 8(b) shows how the sample falls in the
L$_{\lambda5007}$--$N_H$ plane instead.  Although it is less clear cut, it is
generally true that the galaxies with the largest values of L$_{\lambda5007}$
at any given $N_H$ show HBLRs (1925--7245 is the obvious exception here).
Risaliti et al.\ (2000) reported an upper limit for the flux from 0019--7926.
We have compared this limit with other data by ratioing with the extinction
corrected [OIII] flux.  This provides a crude measure of the obscuring column
(eg Bassani et al.\ 1999).  From this we conclude that $N_H>10^{23}$cm$^{-2}$.
We did not plot this result on Figure 8 given the uncertainties involved, but
we note that this limit is consistent with the results from the other non-HBLRs
(though see also the discussion in the Appendix).

At some level the relatively smooth delineation between HBLRs and non-HBLRS in
Figure 8(a) compared to the rougher trends seen in Figure 8(b) implies it is
not simply AGN luminosity, but how that luminosity dominates over the host
galaxy that is important in the detection of HBLRs.  It should be noted that
the separation seen in Figure 8(a) holds true even though there are many lower
limits.  Essentially these data points can only move to the right, and not
vertically.  The separation between the HBLRs and non-HBLRs at high column
density is approximately a horizontal line in this plot, so the HBLRs and
non-HBLRs from our sample can never mix in this diagram.  Of course many of the
non-HBLRs have not been observed in the hard x-ray band.  However, all bar
three of these galaxies have observed W$_{\lambda5007}<$40\AA, and therefore
are consistent with the trends already shown in Figure 8(a) regardless of what
their x-ray properties might be.  The exceptions are Mrk~1361, NGC~5929 and
NGC~7592 which have W$_{\lambda5007}\sim$60\AA.  Hard x-ray observations of
these galaxies would be valuable in determining whether the trends seen hold
for the entire sample, as would spectropolarimetry of NGC~7592.

Finally, we note there is no correlation between $N_H$ and IRAS colour, in
agreement with Alexander (2001).  This confirms the overall suspicion that IRAS
colour is not directly linked to obscuration.

\section{Discussion}
\subsection{Analysis}
It is worth stressing what our spectropolarimetric data actually reveal before
considering how we can explain these observations.  Clearly, the HBLRs, by
definition, show evidence of an obscured (hidden) BLR, consistent with the
predictions of the unified model.  By contrast, we {\em cannot} say that the
non-HBLRs do {\em not} contain a BLR.  All we can say is that for the most part
we do not detect scattered broad H$\alpha$ from such a region at the level we
would predict (see Section 3.1).  For most of the non-HBLRs the signal-to-noise
in our data is sufficient that we can strongly rule our such emission at the
predicted level.  The conclusion must be that any scattered emission from the
BLR, if it exists in these galaxies, appears at a considerably lower level than
our prediction on the basis of the behaviour of the known HBLRs.  For some of
the non-HBLRs we also know the BLR does exist from hard x-ray observations.
Although not relevant to the discussion of why we see an optical HBLR in some
galaxies, this does provide additional confirmation that the unified model is
correct at some level, and indicates that the non detection of a HBLR is not
because there is no BLR to see.  Clearly it would be useful to have x-ray
spectra of the entire sample.  This should now be possible, even for the very
faint Compton thick sources, using Chandra and XMM-Netwon.  

It may also be true that with sufficient signal-to-noise we would discover
broad H$\alpha$ in spectropolarimetry in all Seyfert 2s.  Clearly, our data
shows that for many of the non-HBLRs the scattered flux must be significantly
below our predicted values.  Therefore only data which can actually probe an
order of magnitude deeper in the scattered flux is probably of use.  It may be
that instrumental limitations on the accuracy that can be achieved in the error
in the polarisation will actually prevent such observations from being achieved
even with larger telescopes.  The discussion below allows the possibility that
every galaxy contains an HBLR, since our basic assumption in analysing these
data is that all Seyfert 2s do contain a BLR.  If we can actually detect such
emission in the future, the debate would then centre on whether there are
correlations between the detected scattered broad H$\alpha$ flux and factors
such as AGN luminosity rather than simple detectability.  For the moment,
however, we are limited to seeking an explanation as to why some galaxies show
obvious HBLRs and others show nothing.

There is a relatively simple explanation for all our data, even though our
results may appear rather complex at first sight.  First we note the clear
message from Figure 3 is that HBLRs do appear like reddened Seyfert 1s. This is
evidence that the AGN largely dominates L$_{FIR}$ in these systems at least.
The radio data tend to support this conclusion.  By contrast the non-HBLRs that
lie along the starburst/LINER track clearly cannot have dominant AGN cores.
Therefore the clear split in colour between the bulk of the non-HBLRs and the
HBLRs is due to the relative luminosities of the AGN and the host galaxy as
first proposed by Alexander (2001).  The division between the two groups in the
[OIII] 5007\AA\ to H$\beta$ ratio seen in Figure 6(b) is largely due to the
same cause.  The galaxies with prominent starbursts will have enhanced H$\beta$
relative to [OIII] leading to lower ratios on average in the non-HBLR
population.  Another plausible reason for lower [OIII] 5007\AA\ to H$\beta$
ratios in some of the higher luminosity non-HBLRs is stratification in the NLR.
If the higher excitation lines arise nearer the nucleus, as is often observed
for Seyferts, more obscured objects will naturally have lower [OIII] 5007\AA\
to H$\beta$ ratios, and lower W$_{\lambda5007}$ as well.  This mechanism may
explain the behaviour of some of the Compton thick non-HBLRs for example.
However, both the IRAS colours and the optical line ratios are essentially
secondary indicators of the underlying reason as to why we do or do not see
HBLRs.

The dominant reason we see an HBLR in any galaxy is revealed when we study the
galaxies at all wavelengths at which the AGN dominates.  The HBLRs are (i) more
luminous [OIII] 5007\AA\ sources, and in particular have more prominent Seyfert
characteristics in the optical with large equivalent width line emission; (ii)
have more of their mid infrared flux arising from the region around the AGN;
(iii) on average have brighter radio cores; (iv) have less obscuration than
non-HBLRs of similar luminosity.  The first three of these points indicate that
there are two factors present: first how intrinsically luminous the AGN is, and
second whether this dominates the luminosity compared to the rest of the
emission from the host galaxy.  Figure 8(a) indicates the second of these
factors is as important as the first.  Point (iv) indicates that orientation is
still important (at least if we assume the bulk of the obscuration arises in a
torus).  Clearly, for a galaxy that is tilted further from our line of sight,
there will be greater extinction towards the NLR, and hence the observed
scattered light component will be fainter.  This is certainly consistent with
the results for those non-HBLRs which have good signal-to-noise data, where,
as noted above, the limit on any broad H$\alpha$ present is well below the 
expected value from previously known HBLRs.

The most obvious explanation as to why the AGN luminosity and its contribution
to the bolometric luminosity are so vital is that more luminous AGN cores
support larger extended scattering regions.  Clearly, an extended region is
required in those HBLRs where the obscuring column to the BLR is
$>10^{25}$cm$^{-2}$.  This region of high obscuration must itself be compact
(cf the discussion on NGC1068 in the Appendix and in Risaliti et al.\ 1999).
Our data cannot constrain the size of the scattering region however, though
imaging polarimetry can.  Infrared imaging polarimetry of NGC1068 and the
considerably less luminous Circinus Galaxy (Packham et al.\ 1997, Lumsden et
al.\ 1999, Ruiz et al.\ 2000) clearly shows that the latter has a smaller
scattering region.  Other observable indicators also tend to favour such a
correlation between the size of the scattering region and the AGN luminosity
(see Section 4.2).  If this is correct then we predict that the size of the
scattering region in the less luminous, non-HBLR, Compton thin galaxies must be
small.  

The alternative possibility is that we are simply limited by signal-to-noise in
our polarimetry as claimed by Alexander (2001), and perhaps indicated by our
analysis in Section 3.1 for some of the non-HBLRs.  We made an additional test
to determine whether signal-to-noise is a major factor in HBLR detectability by
examining the x-ray data, since the broad H$\alpha$ flux should scale with the
x-ray flux.  We compared the x-ray fluxes in the Compton thin galaxies to test
whether they showed a difference between the HBLR and non-HBLR systems.  There
is a clear distinction between the two groups in terms of luminosity, with the
HBLRs being more luminous as they are at other wavebands, but none in terms of
detected flux.  This suggests again that luminosity is the key factor rather
than simple signal-to-noise.  It may be the case that those galaxies with a
high current star formation rate also have more molecular gas near the nucleus,
making it easier to hide the NLR.  Further infrared imaging polarimetry, which
has the benefit of penetrating the obscuration whilst minimising any
contribution to the polarization from the host galaxy, may allow us to isolate
the scattering sites even in the less luminous or more heavily obscured systems
and determine if it is the size of the scattering region or simply low
signal-to-noise that best explains our results.

Better spatial and spectral resolution would also be a benefit in future
spectropolarimetric surveys.  Improved spatial resolution will allow the
scattering region to be isolated, reducing the diluting contribution from the
host galaxy, as appears to be the case for Circinus.  This could be
particularly important for those cases where there is significant star
formation.  Young stars are a likely candidate for many galaxies showing
evidence for the so called second featureless continuum (Tran 1995b, and see
also the discussion in Gonzalez Delgado et al.\ 1998).  These stars may also
help mask any polarization signal, since they dilute the direct flux component
from the AGN, resulting in a lower net polarization.  Improved spectral
resolution will make it easier to remove the contribution of polarized narrow
lines, as shown graphically by Young et al.\ (1996a) for the case of
0518--2524.  

The presence of star formation is not in itself an argument against orientation
dependent obscuration in the cooler galaxies however.  At least for AGN of
equivalent luminosity, galaxies with significant star formation in which the
core is obscured are more likely to be characterised as having starburst
activity than those where there is little obscuration.  This may actually be
the case for at least some of our sample, particularly for galaxies with
approximately equal inferred AGN luminosity such as NGC5135, NGC7130, NGC5995
and 0518--2524, all of which show purely Seyfert indicators in the optical
spectra.  We know there must be a scattering region of some kind in the Compton
thick objects NGC5135 and NGC7130, since the hard x-ray continuum from these
objects is {\em reflected} light from the Seyfert 1 core.  Again, we should be
able to test the extent, and the extinction to this scattering region using
infrared imaging polarimetry.  

\subsection{Comparison with other work}
A detailed discussion of other observations from the literature of some of the
galaxies in our sample is presented in the Appendix.

Kay and Moran (1998) and Moran et al.\ (2000) recently reported several new
HBLRs.  Examination of the IRAS fluxes of these galaxies show that they too
have warm colours (with the exception of NGC3081 for which no IRAS data exist).
The same is true of all other previously known HBLRs.  Only three of the
galaxies studied by Moran et al.\ have been observed in the hard x-ray band,
including NGC3081, which has $N_H=6600^{+1800}_{-1600}\times10^{20}$cm$^{-2}$.
Of the other two, NGC2273 has $N_H>10^{25}$cm$^{-2}$ and the coolest colour
with F$_{60\mu{\rm m}}/$F$_{25\mu{\rm m}}=4.4$, and NGC4507 has
$N_H=2920\pm230\times10^{20}$cm$^{-2}$ and F$_{60\mu{\rm m}}/$F$_{25\mu{\rm
m}}=3.1$ (where we have taken the neutral hydrogen column densities from
Risaliti, Maiolino and Salvati 1999).  Although the equivalent widths of [OIII]
5007\AA\ are rarely reported, it is clear from the published spectra in Moran
et al.\ that these galaxies all have W$_{\lambda5007}\gs100$\AA, consistent
with the trends we find.  The same is also true of the high polarization HBLRs
considered by Awaki et al.\ (2000), and those HBLRs found by Young et al.\
(1996a).  In fact the only exception to the trend shown in Figure 8(a) that we
could find in the literature was for Circinus, which has
W$_{\lambda5007}\sim30$\AA, but $N_H=4.3\times10^{25}$cm$^{-2}$ (Matt et al.\
1999).  It may be that the proximity of Circinus (so that we have higher
quality data, with better spatial resolution) is the major factor in enabling
us to see the HBLR.

The case for an extended scattering region in the highest luminosity sources is
also made, in part, by Awaki et al.\ (2000).  They show that almost all of the
highly polarized HBLRs found by Miller and Goodrich (1990) and Tran, Miller and
Kay (1992) are Compton thick, and hence that we see the x-ray emission as well
as the optical through scattered light.  This implies the scattering region
must project outside the area of highest obscuration in all these sources.
Their analysis indicates the x-ray emission can arise up to 30pc away from the
actual core, considerably larger than the putative 1pc scale height of a torus.
Indeed it is possible that these most luminous Seyfert 2s only appear as such
because they are largely edge-on.  There is certainly a dearth of objects in
the top left hand area of Figure 8(a), and of all the objects considered by
Awaki et al.\, only Was 49b lies in this region.  Perhaps such objects are in
fact classified as reddened Seyfert 1s, or 1.5s, instead of Seyfert 2s.
Lastly, we note that the split in x-ray behaviour reported by Awaki et al.\
between the highly polarized HBLRs from the Miller and Goodrich and Tran,
Miller and Kay studies and their sample of `normal' Seyferts is in fact
primarily a split between those systems which are Compton thick and show only
reflected emission from the core, and those in which the core can be seen
through the obscuring material.  The objects they report as non-HBLRs actually
contain several known HBLRs (including four in the sample reported here), and
therefore their comparison of the cause of the HBLR/non-HBLR split should be
viewed instead as the split between Compton thick and Compton thin sources.
Therefore the case they make that HBLRs must all be edge on is
only strictly applicable to the highly polarized sources.

\section{Conclusions}
We have completed a survey of the spectropolarimetric properties of a far
infrared selected sample of Seyfert 2s.  Our sample spans a wide range of 
AGN core luminosity, and appears representative of the local Seyfert
2 population.   We conclude the following from our data:

\begin{list}{$\bullet$}{\topsep=0pt\parsep=0pt\itemsep=0pt\rightmargin=0pt
\leftmargin=0pt\itemindent=2em}
\item{The key factors determining the visibility of an HBLR are (i) AGN
luminosity, and how much the AGN dominates the bolometric luminosity
of the host galaxy and (ii) the level
of obscuration to the AGN.  There is strong evidence that these are
the only substantially important factors (cf Figure 8a).}
\item{We find a similar fraction of HBLRs in our sample as in the optically
selected sample of Moran et al.\ (2000) of 29\%.  It is likely that higher
quality observations, or those at longer wavelengths may help to increase this
fraction.  }
\item{There appears to be a split in the properties of our sample between
galaxies which have a dominant host galaxy component and those which 
have a dominant AGN component as determined by their optical spectral
properties.  There is support for the conclusions of Alexander (2001)
that at least the former group have overall properties that reflect the
the host galaxy rather than the AGN.
There is also support for the opposite point of view for the
AGN dominated sources, that their overall properties are largely 
determined by the AGN luminosity and its obscuration along our line
of sight.}
\item{Our data are still consistent with the unified model, though perhaps not
in its simplest incarnation.  For the subset of data for which both
spectropolarimetry and x-ray spectroscopy exist, all show either scattered
broad H$\alpha$, transmitted Seyfert 1 x-ray spectra or scattered hard x-ray
continua, consistent with a central Seyfert 1 core.  However, our results do
seem to contradict the impression that has grown recently that only obscuration
is important in determining the observational characteristics of these moderate
luminosity AGN (see, eg, Veron-Cetty and Veron 2000).}

\end{list}

\section*{Acknowledgements}
We thank both the anonymous referee and David Alexander for useful suggestions
that improved the presentation of this paper.  We acknowledge awards of
telescope time on the AAT and WHT from ATAC and PATT respectively.  SLL
acknowledges support from PPARC through the award of an Advanced Research
Fellowship.  SLL also thanks the Access to Major Research Facilities Program,
administered by the Australian Nuclear Science and Technology Organisation on
behalf of the Australian Government, for travel support for the observations
reported here.  This research has made use of the NASA/IPAC Extragalactic
Database (NED) which is operated by the Jet Propulsion Laboratory, California
Institute of Technology, under contract with the National Aeronautics and Space
Administration.

\parindent=0pt

\vspace*{3mm}

{\bf References}\par
\mnref{Alexander, D.M., 2001, \mn, 320, L15}
\mnref{Alexander, D.M., Aussel, H., 2000, preprint}
\mnref{Antonucci, R. 1993, ARA\&A, 31, 473}
\mnref{Antonucci, R., Miller, J.S. 1985, ApJ, 297, 621}
\mnref{Arextaga, I, Joguet, B., Kunth, D., Melnick, J., Terlevich, R.J., 
	1999, \apj, L123}
\mnref{Awaki, H., Ueno, S., Taniguchi, Y., Weaver, K.A., 2000, \apj,  
	542, 175} 
\mnref{Bassani, L., Dadina, M., Maiolino, R., Salvati, M., Risaliti, G., 
	della Ceca, R., Matt, G., Zamorani, G., 1999, \apjs,  121, 473}
\mnref{Bicay, M.D., Kojoian, G., Seal, J., Dickinson, D.F., Malkan, M.A.,
	1995, \apjs, 98, 369}
\mnref{Bransford, M.A., Appleton, P.N., Heisler, C.A., Norris, R.P., 
	Marston, A.P., 1998, \apj, 497, 133}
\mnref{Capetti, A., Axon, D.J., Macchetto, F., Sparks, W.B.,
	Boksenberg, A., 1995, \apj,  446, 155} 
\mnref{Capetti, A., Axon, D.J., Macchetto, F.D., 1997, \apj, 487, 560} 
\mnref{Chini, R., Kreysa, E., Biermann, P.L., 1989, \aaa, 219, 87}
\mnref{Clavel, J. et al., 2000, \aaa,  357, 839} 
\mnref{Colina, L., Sparks, W.B., Macchetto, F., 1991, \apj, 370, 102}
\mnref{Colina, L., Lipari, S., Macchetto, F., 1991, \apj,  379, 113} 
\mnref{Condon, J.J., Helou, G., Sanders, D.B., Soifer, B.T., 1990, 
	\apjs, 73, 359} 
\mnref{Condon, J.J., Anderson, E., Broderick, J.J., 1995, \aj, 109, 2318} 
\mnref{Condon, J.J., Cotton, W.D., Greisen, E.W., Yin, Q.F., Perley, R.A.,
	 Taylor, G.B., Broderick, J.J., 1998, AJ, 115, 1693}
\mnref{Coziol, R., Demers, S., Pena, M., Torres-Peimbert, S., Fontaine, G., 
	Wesemael, F.,  Lamontagne, R., 1993, \aj, 105, 35} 
\mnref{Dahari, O., 1985, \apjs, 57, 643}
\mnref{de Grijp, M.H.K., Keel, W.C., Miley, G.K., Goudfrooij, P., Lub, J., 
	1992, \aas, 96, 389} 
\mnref{de Jong, T., Klein, U., Wielebinski, R., \& Wunderlich, E., 1985, 
	\aaa, 147, L6} 
\mnref{De Robertis, M.M., Hutchings, J.B., Pitts, R.E., 1988, \aj, 95, 1371}
\mnref{Devereux, N., 1987, \apj,  323, 91} 
\mnref{Dopita, M.A., Heisler, C.A., Lumsden, S.L., Bailey, J.A., 1998, \apj,
	498, 570}
\mnref{Efstathiou, A., Rowan-Robinson, M., 1995, \mn, 273, 649}
\mnref{Feigelson, E.D., Nelson, P.I., 1985, \apj,  293, 192} 
\mnref{Gallimore, J.F., Baum, S.A., O'Dea, C.P., 1997, Nature, 388, 852} 
\mnref{Giuricin, G., Mardirossian, F., Mezzetti, M.,  Bertotti, G., 1990, 
	\apjs, 72, 551} 
\mnref{Giuricin, G., Mardirossian, F., Mezzetti, M., 1995, \apj,  446, 
	550} 
\mnref{Gonz{\'a}lez Delgado, R.M., Heckman, T., Leitherer, C., Meurer, G., 
	Krolik, J., Wilson, A.S., Kinney, A., Koratkar, A., 1998, \apj,  
	505, 174} 
\mnref{Gonz{\'a}lez Delgado, R.M., Heckman, T., Leitherer, C., 2001, \apj, 
	546, 845}
\mnref{Granato, G.L., Danese, L., 1994, \mn, 268, 235}
\mnref{Guainazzi, M., Matt, G., Antonelli, L.A., Fiore, F., Piro, L.,
	Ueno, S., 1998, \mn,  298, 824} 
\mnref{Guainazzi, M., Molendi, S., Vignati, P., Matt, G., Iwasawa, K., 
	2000, New Astronomy,  5, 235} 
\mnref{Heisler, C.A., Lumsden, S.L., Bailey, J.A., 1997, Nature, 385, 700}
\mnref{Heisler, C.A., Norris, R.P., Jauncey, D.L., Reynolds, J.E., 
	King, E.A., 1998, \mn, 300, 1111}
\mnref{Helou, G., Soifer, B.T., Rowan-Robinson, M., 1985, \apj, 298, 
	L7} 
\mnref{Ho, L.C., Filippenko, A.V., Sargent, W.L.W., 1997, \apjs, 112, 315}
\mnref{Huchra, J.P., Burg, R., 1992, \apj, 393, 90}
\mnref{Inglis, M.D., Brindle, C., Hough, J.H., Young, S., Axon, D.J., Bailey,
	J.A., Ward, M.J., 1993, \mn, 263, 895}
\mnref{Inglis, M.D., Young, S., Hough, J.H., Gledhill, T., Axon, D.J., Bailey,
	J.A., Ward, M.J., 1995, \mn, 275, 398}
\mnref{IRAS Point Source Catalog, Version 2. 1988, Joint IRAS Science Working
	Group, US GPO, Washington DC}
\mnref{Isasawa, K., Fabian, A.C, Brandt, C.S., Crawford, C.S., Almaini, O.,
	1997, \mn, 291, L17}
\mnref{Isobe, T., Feigelson, E.D., Nelson, P.I., 1986, \apj,  306, 
	490} 
\mnref{Kay, L.E., 1994, \apj, 430, 196}
\mnref{Kay, L.E., Moran, E.C., 1998, PASP,  110, 1003} 
\mnref{Keel, W.C., Kennicutt, R.C., Hummel, E., van der Hulst, J.M., 
	1985, \aj, 90, 708} 
\mnref{Kewley, L.J., Heisler, C.A., Dopita, M.A., Sutherland, R., 
	Norris, R.P., Reynolds, J., Lumsden, S.L., 2000, \apj, 530, 704}
\mnref{Kewley, L.J., Heisler, C.A., Dopita, M.A., Lumsden, S., 2001, \aj,
	       132, 37}
\mnref{Kii, T., Nakagawa, T., Fujimoto, R., Ogasaka, T., Miyazaki, T., Kawabe,
R., Terashima, Y., 1996, in X-Ray Imaging and Spectroscopy of Hot Cosmic
Plasmas, ed.\ F.\ Makino \& K.\ Mitsuda (Tokyo: Universal Academy Press), 161}
\mnref{Laor, A., Draine, B.T., 1993, \apj, 402, 441}
\mnref{LaValley, M., Isobe, T., Feigelson, E.D., 1992, BAAS, 24, 839}
\mnref{Lonsdale, C.J., Lonsdale, C.J., Smith, H.E., 1992, \apj, 
	391, 629} 
\mnref{Lumsden, S.L., Moore, T.J.T., Smith, C., Fujiyoshi, T., 
	Bland-Hawthorn, J., Ward, M.J., 1999, \mn,  303, 209} 
\mnref{Maiolino, R., Ruiz, M., Rieke, G.H., Keller, L.D., 1995, 
	\apj,  446, 561} 
\mnref{Malaguti, G. et al., 1998, \aaa,  331, 519} 
\mnref{Matt, G., et al., 1997, \aaa, 325, L13}
\mnref{Matt, G., et al., 1999, \aaa,  341, L39} 
\mnref{Matt, G., Fabian, A.C., Guainazzi, M., Iwasawa, K., Bassani, L.,
	Malaguti, G., 2000, \mn,  318, 173} 
\mnref{Miller, J.S., Mathews, W.G., 1972, \apj, 172, 593}
\mnref{Miller, J.S., Robinson, L.B., Goodrich, R.W., 1988, in Instrumentation
	for Ground Based Astronomy, p 157, ed. L.B. Robinson, Springer,
	New York}
\mnref{Miller, J.S., Goodrich, R.W., 1990, \apj, 355, 456}
\mnref{Miller, J.S., Goodrich, R.W., Mathews, W.G., 1991, \apj, 378, 47}
\mnref{Moran, E.C., Halpern, J.P., Helfand, D.J., 1996, \apjs, 106, 341} 
\mnref{Moran, E.C., Barth, A.J., Kay, L.E., Filippenko, A.V., 2000, \apj,
	540, L73}
\mnref{Morganti, R., Tsvetanov, Z.I., Gallimore, J., Allen, M.G., 
	1999, \aas, 137, 457} 
\mnref{Moshir, M., et al., 1991, Explanatory Supplement to the IRAS Faint
	Source Survey, Version 2.  JPL, Pasadena}
\mnref{Muxlow, T.W.B., Pedlar, A., Holloway, A.J., Gallimore, J.F., 
	Antonucci, R.R.J., 1996, \mn,  278, 854} 
\mnref{Norris, R.P., Allen, D.A., Roche, P.F., 1988, \mn, 234, 773} 
\mnref{Osterbrock, D.E. 1993, \apj, 404, 551}
\mnref{Osterbrock, D.E., Martel, A., 1993, \apj, 414, 552}
\mnref{Packham, C., Young, S., Hough, J.H., Axon, D.J., Bailey, J.A., 
	1997, \mn,  288, 375} 
\mnref{Pappa, A., Georgantopoulos, I., Stewart G.C., 2000, \mn, 314, 589}
\mnref{P{\'e}rez Garc{\'i}a, A. M., Rodr{\'i}guez Espinosa, J.M.,
	2000, preprint} 
\mnref{Pier, A., Krolik, J., 1992, \apj, 401, 99}
\mnref{Risaliti, G., Maiolino, R., Salvati, M., 1999, \apj, 522, 157}
\mnref{Roy, A.L., Norris, R.P., 1997, \mn, 289, 824}
\mnref{Roy, A.L., Norris, R.P., Kesteven, M.J., Troup, E.R., Reynolds, J.E., 
	1994, \apj, 432, 496}
\mnref{Roy, A.L., Norris, R.P., Kesteven, M.J., Troup, E.R., Reynolds, J.E., 
	1998, \mn, 301, 1019}
\mnref{Ruiz, M., Alexander, D.M., Young, S., Hough, J., Lumsden, S.L., 
	Heisler, C.A., 2000, \mn,  316, 49} 
\mnref{Rush, B., Malkan, M.A., Spinoglio, L., 1993, \apjs, 89, 1}
\mnref{Rush, B., Malkan, M.A., Edelson, R.A., 1996, \apj, 473, 130}
\mnref{Ruiz, M., Rieke, G.H., Schmidt, G.D., 1994, \apj, 423, 608}
\mnref{Sadler, E.M., Slee, O.B., Reynolds, J.E., Roy, A.L., 1995, 
	\mn, 276, 1373}
\mnref{Sanders, D.B., Egami, E., Lipari, S., Mirabel, I.F., Soifer, B.T.,
	1995, \aj, 110, 1993}
\mnref{Schachter, J.F., Fiore, F., Elvis, M., Mathur, S., Wilson, A.S., 
	Morse, J.A., Awaki, H., Iwasawa, K., 1998, \apj,  503, L123}
\mnref{Schinnerer, E., Eckart, A., Tacconi, L.J., 1999, \apj,  524, 
	L5} 
\mnref{Shields, J.C., Filippenko, A.V., 1996, \aaa,  311, 393} 
\mnref{Soifer, B.T., Boehmer, L., Neugebauer, G., Sanders, D.B., 1989,
	\aj, 98, 766}
\mnref{Sopp, H.M., Alexander, P., 1991, \mn, 251, 14P} 
\mnref{Sopp, H., Alexander, P., 1992, \mn, 259, 425}
\mnref{Stauffer, J.R., 1982, \apj, 262, 66} 
\mnref{Stirpe, G.M., 1990, \aas, 85, 1049}
\mnref{Strauss, M.A.,  Huchra, J.P., Davis, M., Yahil, A., Fisher, K.B.,
	 Tonry, J., 1992, \apjs, 83, 29}
\mnref{Su, B.M., Muxlow, T.W.B., Pedlar, A., Holloway, A.J., Steffen, W.,
	Kukula, M.J., Mutel, R.L., 1996, \mn, 279, 1111}
\mnref{Telesco, C.M., Becklin, E.E., Wynn-Williams, C.G., Harper, D.A., 
	1984, \apj,  282, 427}
\mnref{Terashima, Y., Ptak, A., Fujimoto, R., Itoh, M., Kunieda, H., 
	Makishima, K., Serlemitsos, P.J., 1998, \apj,  496, 210} 
\mnref{Thean, A., Pedlar, A., Kukula, M.J., Baum, S.A., O'Dea, C.P., 
	2000, \mn, 314, 573}
\mnref{Tran, H.D., 1995a, \apj, 440, 565}
\mnref{Tran, H.D., 1995b, \apj, 440, 597}
\mnref{Tran, H.D., Miller, J.S., Kay, L.E., 1992, \apj, 397, 452}
\mnref{Tran, H.D., Brotherton, M.S., Stanford, S.A., van Breugel, W., Dey, A., 
	Stern, D., Antonucci, R., 1999, \apj, 516, 85}
\mnref{Turner, T.J., George, I.M., Nandra, K., Mushotzky, R.F., 1997a, 
	\apj, 488, 164}
\mnref{Turner, T.J., George, I.M., Nandra, K., Mushotzky, R.F., 1997b, 
	\apjs, 113, 23}
\mnref{Turner, T.J., Perola, G.C., Fiore F., Matt G., George, I. M., 
	Piro L., Bassani L., 2000, \apj,  531, 245}
\mnref{Ueno, S., 1997, PhD thesis, Kyoto Univ.}
\mnref{Ulvestad, J.S., Wilson, A.S., 1989, \apj, 343, 659}
\mnref{Vader, J.P., Frogel, J.A., Terndrup, D.M., Heisler, C.A., 
	1993, \aj, 106, 1743} 
\mnref{Vacca, W.D., Conti, P.S., 1992, \apj, 401, 543}
\mnref{Veilleux, S., Osterbrock, D.E., 1987, \apjs, 63, 295}
\mnref{Veilleux, S., Kim, D.-.C, Sanders, D.B., Mazzarella, J.M.,
	Soifer, B.T., 1995, \apjs,  98, 171} 
\mnref{V{\'e}ron, P., Goncalves, A.C., V{\'e}ron-Cetty, M.P., 1997, \aaa, 
	319, 52} 
\mnref{V{\'e}ron-Cetty, M.P., V{\'e}ron, P., 2000, \aaa R,  10, 81} 
\mnref{Xue, S., Otani, C., Mihara, T., Cappi, M., Matsuoka, M., 1998, 
	PASJ,  50, 519 }
\mnref{Young, S., 2000, \mn,  312, 567}
\mnref{Young, S., Hough, J.H., Axon, D.J., Bailey, J.A., Ward, M.J., 
	1995, \mn,  272, 513} 
\mnref{Young, S., Hough, J.H., Efstathiou, A., Wills, B.J., Bailey, J.A.,
	Ward, M.J., Axon, D.J., 1996a, \mn, 281, 1206}
\mnref{Young, S., Packham, C., Hough, J.H., Efstathiou, A., 1996b, 
	\mn, 283, L1}

\section*{Appendix: Comments on Individual Sources}
Many of the galaxies discussed in this paper have been the subject of other
detailed individual studies.  It is worth highlighting some aspects of these
studies since they illuminate the possible underlying physical conditions.

{\em 0019--7926:} As noted in Section 2.5 both de Grijp et al.\ (1992) and
Vader et al.\ (1993) present spectra of this object showing clearer Seyfert
characteristics.  From the data in Vader et al.\ we estimate that
W$_{\lambda5007}\sim200$\AA, and the intrinsic line luminosity is $\sim6$ times
larger than that reported here.  This would move 0019--7926 in Figure 8 from
the non-HBLR region into the HBLR one.  We therefore suspect that new
spectropolarimetric observations may detect an HBLR in this system.  If true,
this would increase the distinction between the HBLRs and non-HBLRs in terms of
L$_{\lambda5007}$, but not significantly change our other conclusions.

{\em NGC1068:} This is the best studied of all the HBLRs currently known.  We
now know that the likely orientation of the parsec scale torus in this galaxy
is nearly edge on from direct radio imaging (Gallimore, Baum and O'Dea 1997).
Unfortunately, the structure of the inner regions of NGC1068 is actually rather
complex, since there is a small parsec scale radio structure aligned with the
axis of this torus that is also visible in [OIII] images and imaging
polarimetry from HST (Muxlow et al.\ 1996, Capetti et al.\ 1995, Capetti, Axon
and Macchetto 1997), but the larger scale radio structure and visible
ionisation cones are rotated through $\sim30^\circ$ relative to these (Capetti,
Axon and Macchetto 1997).

This complexity may however help to explain the variant requirements of the
x-ray and the optical/infrared data.  The BeppoSAX data reported by Guainazzi
et al.\ (2000) show evidence for considerable variability in the hard x-ray
emission from NGC1068.  They use this to constrain the location of the
scattering medium at $\sim$1pc from the nucleus.  Since the Fe K$\alpha$ line
is fed by resonance scattering from a largely neutral medium, it further
implies the obscuring torus is also $\sim$1pc from the nucleus.  Indeed, as has
previously been noted (Risaliti et al.\ 1999), unless the bulk of the
obscuration lies at small radii, the implied mass of molecular gas would exceed
the observed virial mass in NGC1068 and other nearby AGN.  However, there is
also clearly scattering on larger scales as detected both in the near infrared
by Packham et al.\ (1997) and Lumsden et al.\ (1999), and in the ultraviolet
regime by Capetti et al.\ (1995).  Young et al.\ (1996b) found that a large
scale torus ($\sim200$pc) is required to explain the infrared data, whilst
still requiring the inner scattering radius to be $\sim1$pc to explain the
optical spectropolarimetry (Young et al.\ 1995).  The solution to this apparent
dichotomy lies in the fact that there are clearly two levels of obscuring
source in NGC1068.  The mid infrared polarimetry of Lumsden et al.\ (1999)
requires a compact optically thick (A$_V\sim100$) region around the BLR, with a
more diffuse larger scale molecular cloud with A$_V=30$ hiding that region and
the inner scattering cones.  This latter region is also revealed in maps of the
molecular gas near the nucleus (Schinnerer, Eckart, and Tacconi 1999).  Clearly
this also implies however that the reflected x-ray component should itself be
partially absorbed.  In the case of NGC1068, if our estimate of the visual
extinction is correct, and the conversion to a column density is $\sim1/10$
that of the Galactic ISM (which is typical of AGN), then the implied
$N_H\sim5\times10^{21}$cm$^{-2}$.  This is sufficiently small to remain
unnoticed in the existing data.  NGC1068 should be taken as a cautionary
example that local conditions can play a large role in the final global
spectral properties of a galaxy.

Finally, we note that the circumnuclear starburst that lies at a radius of
$\sim1$kpc from the nucleus may actually dominate the far infrared flux from
NGC1068 (Telesco et al.\ 1984).  This suggests that the warmth of the infrared
colours is in fact due to the dominance of the AGN component at shorter
wavelengths.

{\em 0518--2524:} The higher resolution optical spectrum we used in determining
the line ratios in Table 2 (which will be discussed in greater detail in a
paper in preparation) also shows that 0518--2524 has a strong Balmer absorption
line spectrum in the blue.  This is probably also the reason why many previous
papers fail to detect the H$\beta$ emission present.  There is therefore clear
evidence for a recent starburst episode in 0518--2524.  In addition, Clavel et
al.\ (2000) show that there is relatively strong PAH emission in the mid
infrared spectrum of this object, which is usually taken as an indication of
the 10$\mu$m emission being due in significant part to star formation.  It is a
moot point as to whether the large far infrared luminosity is due to any star
formation or the obvious AGN activity.  The relatively low W$_{\lambda5007}$
and failure to detect a compact core at longer radio wavelengths
when a prominent core exists at shorter wavelengths may indicate
the former however.  It is also notable that this galaxy has the coolest
mid-far infrared colours of any of the HBLRs presented here, even though
the inferred extinction to the core from the x-ray data is small, which
also suggests there may be a significant contribution to L$_{FIR}$
from something other than the AGN.

{\em NGC4388:} This galaxy is almost edge-on, so any obscuration hiding the BLR
may actually be due in part to the host galaxy rather than a nuclear torus.
Weak broad H$\alpha$ has been detected in direct light in off nuclear positions
by Shields and Filippenko (1996).  They find that this is best explained as 
evidence for an HBLR, even though the direct evidence from polarimetry
is actually marginal (Young et al.\ 1996a).  The combination of both
observations however gives greater confidence in the classification of this
galaxy as an HBLR.

{\em IC3639:} This galaxy has been studied in detail in the optical and
ultraviolet by Gonzalez Delgado et al.\ (1998) and Gonzalez Delgado, Heckman and
Leitherer (2001).  They find evidence for a recent burst of star formation in
the nucleus which contributes an equal amount to the AGN component to the
overall bolometric luminosity.  The starburst does not dominate the optical
continuum however, since an older population contributes a larger fraction of
the light.

{\em NGC5135 and NGC7130:} Detailed optical and ultraviolet studies of 
these galaxies 
were carried out by Gonzalez Delgado et al.\ (1998) and Gonzalez Delgado,
Heckman and Leitherer (2001).  They found evidence that the optical continuum
was dominated by a young starburst.  As with IC3639, the starburst has a
similar bolometric luminosity to the AGN.

{\em NGC7582:} This galaxy showed brief evidence of broad H$\alpha$ in direct
flux (Aretxaga et al.\ 1999), probably due to a lowering of the obscuring
column density, since there is also evidence for considerable hard x-ray
variability in this galaxy (Schachter et al.\ 1998; Xue et al.\ 1998).  Turner
et al.\ (2000) summarise the evidence for a `patchy' torus in this source, and
the alternative possibility that the broad permitted line arose in a supernova
as suggested by Aretxaga et al.  Turner et al.\ find that the best fit to their
2-100 keV BeppoSAX data requires an obscuring column of
$1.4\times10^{23}$cm$^{-2}$ which completely covers the source, plus a Compton
thick absorber with $N_H\sim1.6\times10^{24}$cm$^{-2}$ which only covers 60\%
of it.  Because the Compton thick component does not cover the entire source,
`holes' in the thinner component could lead to unobscured sight lines towards
the BLR.  The optical classification (Table 2) is a reflection of fact that
there is significant ongoing star formation in this galaxy.

{\em The galaxies without spectropolarimetry:} Four of the galaxies in our
initial sample do not have spectropolarimetry.  There is little available data
for two of these (NGC5427 and NGC5899) but they would appear to be typical of
the other galaxies which have cool IRAS colours in our sample.  NGC7592 (also
known as Mrk928) has been the subject of considerable study.  It consists of an
interacting pair: the eastern component is classed as a pure starburst
(Lonsdale, Lonsdale \& Smith 1992).  The western component is classed as a
possible AGN (see Table 2), though clearly has characteristics of a transition
or composite object.  Veron, Goncalves, \& Veron-Cetty (1997) discuss this
aspect of the galaxy in more detail.  It seems unlikely that any of these
three galaxies will show evidence for an HBLR.  The final `unobserved' galaxy
is 0031--2142.  This system is a known luminous x-ray source (see Moran,
Halpern \& Helfand 1996), but has the appearance of a starburst galaxy from its
optical spectrum.  Only the presence of weak broad H$\alpha$ led to
its classification as a Seyfert.  The recent x-ray study of Georgantopoulos
(2000) has confirmed its identification as a marginally obscured Seyfert.
We have shown where this galaxy falls on our $N_H$--W$_{\lambda5007}$
diagnostic plot.  On the basis of this evidence it is possible that despite
its cool colours this galaxy may show scattered broad H$\alpha$.  Although
the optical and FIR emission is dominated by the starburst component, the
obscuration is low.  In this sense it might be thought of as analogous
to a system such as NGC5995, but with added star formation present.
Spectropolarimetry of this source is certainly desirable.

\onecolumn

\begin{table}
\tabcolsep=3.0pt
\begin{tabular}{lccrrrrrcc}
\multicolumn{1}{c}{Name} &
\multicolumn{1}{c}{RA} &
\multicolumn{1}{c}{Dec } &
\multicolumn{1}{c}{$cz$} &
\multicolumn{1}{c}{F$_{12\mu{\rm m}}$ }
 & \multicolumn{1}{c}{F$_{25\mu{\rm m}}$  }
& \multicolumn{1}{c}{F$_{60\mu{\rm m}}$   }&
\multicolumn{1}{c}{F$_{100\mu{\rm m}}$  }&
\multicolumn{1}{c}{L$_{FIR}$}&
\multicolumn{1}{c}{Flag}
\\
 & (1950) & (1950) & \multicolumn{1}{c}{(km/s)} &
\multicolumn{1}{c}{(Jy)} &
\multicolumn{1}{c}{(Jy)} &
\multicolumn{1}{c}{(Jy)} &
\multicolumn{1}{c}{(Jy)} & 
\multicolumn{1}{c}{(log(\Lsolar))}\\
Mrk334 & 00 00 35.1 & $+$21 40 52 & 6824 & 0.26$\pm0.039$ &1.05$\pm0.095$ &
4.35$\pm0.305$ & 4.32$\pm0.346$ & 10.70 \\
0019$-$7926 &  00 19 52.2 & $-$79 26 46 &21502 & 0.36$\pm0.035$ & 1.27$\pm0.055$ & 3.16$\pm0.134$& 2.94$\pm0.151$& 11.58\\
0031$-$2142 & 00 31 43.1 & $-$21 42 52 & 8133 & 0.22$\pm0.029$ &
0.56$\pm$0.045 & 3.85$\pm0.193$ & 8.42$\pm0.505$ & 10.93 & *\\
NGC1068 & 02 40 07.2 & $-$00 13 29 & 1178 &36.10$\pm0.064$ &  84.25$\pm0.191$ &
181.95$\pm0.103$ & 235.87$\pm0.218$ & 10.83  \\
NGC1143 & 02 52 38.8 & $-$00 23 07 & 8654 & 0.27$\pm$0.030 & 0.58$\pm0.028$ &
5.06$\pm$0.058 & 11.45$\pm$0.198 & 11.12 \\
0425$-$0440 & 04 25 56.9 & $-$04 40 25 &4592 & 0.16$\pm0.022$& 1.43$\pm0.058$&
4.13$\pm0.162$& 3.30$\pm0.192$& 10.32 \\
0518$-$2524 & 05 18 58.6 & $-$25 24 39 &12567 & 0.74$\pm0.021$ &
3.50$\pm0.025$ & 13.95$\pm0.031$ & 12.52$\pm0.081$ & 11.73 \\
NGC4388 & 12 23 14.4 & $+$12 56 22 &2455 & 1.06$\pm0.031$ & 3.42$\pm0.068$ &
10.05$\pm0.033$ & 17.40$\pm0.177$ & 10.26  \\
IC3639 & 12 38 10.2 & $-$36 28 51 & 3059 & 0.64$\pm0.034$ & 2.55$\pm0.044$  &
7.53$\pm0.042$ & 11.54$\pm0.204$ & 10.31 \\
MCG-3-34-64 & 13 19 42.8 & $-$16 27 55 &4999 & 0.93$\pm0.040$ & 2.95$\pm0.041$
& 6.07$\pm0.039$ & 5.93$\pm0.137$ & 10.58  \\
NGC5135 & 13 22 56.7 & $-$29 34 25 &3923 & 0.62$\pm0.038$ & 2.53$\pm0.053$  &
17.10$\pm0.060$ & 29.50$\pm0.179$ & 10.90 \\
NGC5194 & 13 27 45.3 & $+$47 27 24 &579 & 11.00$\pm0.100$ & 15.00$\pm0.150$ &
98.80$\pm0.988$ & 280.40$\pm2.804$ & 10.10  \\
NGC5256 & 13 36 14.1 & $+$48 31 52 & 8353 & 0.28$\pm$0.026 & 1.13$\pm0.037$ &
7.19$\pm0.032$ & 10.35$\pm0.163$ & 11.18\\
Mrk1361 & 13 44 36.5 & $+$11 21 15 &6747 & 0.17$\pm0.039$& 0.84$\pm0.101$& 3.28$\pm0.394$& 3.73$\pm0.373$& 10.63\\
NGC5427 & 14 00 48.3 & $-$05 47 25 & 2647 & 1.14$\pm0.043$ & 1.33$\pm0.070$ &
9.93$\pm0.056$ & 24.81$\pm0.120$ & 10.40 & *\\
1408+1347 & 14 08 16.6 & $+$13 47 32 &4837 & 0.13$\pm0.023$& 1.04$\pm0.073$&
3.69$\pm0.258$& 2.87$\pm0.201$& 10.31 \\
NGC5899 & 15 13 15.0 & $+$42 14 06 & 2706 & 0.52$\pm0.042$ & 0.51$\pm0.041$ &
4.13$\pm0.207$ & 11.43$\pm0.572$ & 10.05 & *\\
NGC5929 & 15 24 20.6 & $+$41 50 57 &2708 & 0.43$\pm0.033$ & 1.62$\pm0.026$ &
9.14$\pm0.041$ & 13.69$\pm0.124$ & 10.28 \\
NGC5995 & 15 45 37.4 & $-$13 36 17 &7297 & 0.53$\pm0.041$& 1.40$\pm0.098$& 3.95$\pm0.277$& 6.70$\pm0.536$& 10.82\\
1925$-$7245 & 19 25 27.8 & $-$72 45 39 &18339 & 0.23$\pm0.020$ & 1.34$\pm0.024$
& 5.30$\pm0.031$ & 6.70$\pm0.165$ & 11.68 \\
IC5063 & 20 48 11.7 & $-$57 15 26 &3321 & 1.16$\pm0.025$ & 4.00$\pm0.033$ &
6.11$\pm0.041$ & 4.31$\pm0.206$ & 10.19 \\
NGC7130 &21 45 19.7 & $-$35 11 03 &4875 &0.62$\pm0.020$ & 2.18$\pm0.029$ & 16.91$\pm0.045$ & 25.97$\pm0.161$ & 11.06\\
NGC7172 & 21 59 07.0 & $-$32 06 42 &2622 & 0.49$\pm0.029$ & 0.94$\pm0.008$ &
5.99$\pm0.026$ & 12.02$\pm0.107$ & 10.12 \\
NGC7479 & 23 02 26.6 & $+$12 03 08 & 2586 & 1.40$\pm$0.037 & 3.92$\pm$0.066 &
15.35$\pm$0.060 & 24.60$\pm$0.308 & 10.49 \\
IC5298 & 23 13 31.2 & $+$25 16 48 &8449 & 0.32$\pm0.033$ & 1.88$\pm0.061$ &
8.75$\pm0.052$ & 11.64$\pm0.123$ & 11.23 \\
NGC7582 & 23 15 38.1 & $-$42 38 40 &1552 &  2.30$\pm0.030$ & 7.49$\pm0.028$ & 51.64$\pm0.108$ & 78.36$\pm0.121$ & 10.55\\
NGC7592 & 23 15 47.5 & $-$04 41 20 & 7480 & 0.27$\pm0.034$ & 0.95$\pm0.067$ &
8.02$\pm0.053$ & 10.50$\pm0.149$ & 11.09 & *\\
NGC7674 & 23 25 24.7 & $+$08 30 14 &8895 & 0.68$\pm0.043$ & 1.88$\pm0.044$ & 5.28$\pm0.048$ & 7.91$\pm0.171$ & 11.08\\
\end{tabular}

\caption{Complete IRAS selected sample for spectropolarimetry observations.
Redshift data are taken from Strauss et al.\ (1992).  IRAS fluxes are derived
from the IRAS BGS survey (Soifer et al.\ 1989; Sanders et al.\ 1995) except for
0019$-$7926, NGC5995 and NGC5899 which are from Strauss et al.\ (1990), and
Mrk334, 0031$-$2142, 0425$-$0440, Mrk1361 and 1408$+$1347 which were taken from
FSC.  Galaxies which have a $*$ in the flag column do not have
spectropolarimetry.
 }

\end{table}

\begin{table}
\tabcolsep=3.0pt
\begin{tabular}{lcccrrrrrrrrccccr}
\multicolumn{1}{c}{Name} &
\multicolumn{1}{c}{HBLR?} &
\multicolumn{1}{c}{Ref} &
\multicolumn{1}{c}{E(B--V)} &
\multicolumn{1}{c}{${\frac{\rm [OIII]}{{\rm H}\beta}}$} &
\multicolumn{1}{c}{${\frac{\rm [OI]}{{\rm H}\alpha}}$} &
\multicolumn{1}{c}{${\frac{\rm [NII]}{{\rm H}\alpha}}$} &
\multicolumn{1}{c}{${\frac{\rm [SII]}{{\rm H}\alpha}}$} &
\multicolumn{1}{c}{${\frac{\rm HeII}{{\rm H}\beta}}$} &
\multicolumn{1}{c}{W$_{\lambda5007}$} &
\multicolumn{1}{c}{F$_{\lambda5007}$} &
\multicolumn{4}{c}{Classification}
&\multicolumn{1}{c}{Ref}
\\
\multicolumn{11}{c}{ } &
\multicolumn{1}{c}{(1)}&\multicolumn{1}{c}{(2)}&
\multicolumn{1}{c}{(3)}&\multicolumn{1}{c}{(4)}\\
Mrk334      &n?&b  & 0.69&  0.23& $-$1.28& $-$0.23& $-$0.55&  0.00&
27.0&2.00&L&I&I&N&1 2  \\
0019$-$7926 &n\phantom{?}&a  & 0.84&  0.47& $-$1.36& $-$0.31& $-$0.51& $-$1.18& 30.7&0.13&I&I&I&S& 3 \\
0031$-$2142 & & & 0.44 & $-$0.03 & $-$1.30 & $-$0.35 & $-$0.67 & & 10.0 & 0.12
& I & H & I &  & 4 5\\
NGC1068     &y\phantom{?}&cd & 1.02&  1.07& $-$0.87&  0.24& $-$0.53&
$-$0.43&460.0&47.00&S&S&S&S&3 4 \\
NGC1143 & n\phantom{?} & a & 0.75 & 1.17 & $-$0.70 & 0.33 & $-$0.04 & $-$0.76 & 34.7 &
0.48 & S & S & S & S & 3\\
0425$-$0440 &n?&a  & 1.93&  0.81& $-$0.69& $-$0.05& $-$0.41&  $<-0.36$&  6.6&1.30&S&S&S&N& 3 \\
0518$-$2524 &y\phantom{?}&e  & 0.32&  1.05& $-$0.55&  0.30& $-$0.23& $-$0.26& 45.3&1.30&S&S&S&S& 3 \\
NGC4388     &y\phantom{?}&e  & 0.61&  1.05& $-$0.80& $-$0.24& $-$0.21&
$-$0.62&908.0&2.20&S&S&S&S&7 8 \\
IC3639      &y\phantom{?}&a  & 0.85&  0.91& $-$0.82& $-$0.09& $-$0.37& $-$0.78&110.0&3.40&S&S&S&S& 3 \\
MCG-3-34-64 &y\phantom{?}&e  & 0.30&  1.06& $-$0.64&  0.10& $-$0.38& $-$0.41&191.0&4.00&S&S&S&S&9 \\
NGC5135     &n\phantom{?}&a  & 0.78&  0.82& $-$1.25& $-$0.09& $-$0.54& $-$0.64& 82.7&2.30&S&S&S&S& 3 \\
NGC5194     &n?&a  & 0.97&  1.05& $-$0.74&  0.51& $-$0.03&  $<-0.51$& 11.0&2.20&S&S&S&N& 3 \\
NGC5256 & n\phantom{?} & a   & 0.51 & 0.67 & $-$1.20 & $-$0.24 & $-$0.47 & $-$0.69 &
101.1 & 0.21 & S & S & S & S & 3\\
Mrk1361     &n\phantom{?}&a  & 1.08&  0.70& $-$1.10& $-$0.14& $-$0.72& $-$0.90& 80.0&1.80&S&I&S&S& 3 \\
NGC5427     & &   & 0.00 & 0.91 & $-$0.57 & 0.21 & $-$0.29 && 35.0 & 0.09 & 
S&S&S&& 10\\
1408+1347   &?\phantom{?}&a  & 1.04&  0.67& $-$0.77&  0.11& $-$0.21&  $<-1.09$& 24.3&0.08&S&S&S&N& 3 \\
NGC5899     & &   & 0.58 & 1.02 & $-$0.64 & 0.32 & $-$0.05 & &45.0 & 0.41 &
S&S&S&& 11\\
NGC5929     &n\phantom{?}&a  & 0.37&  0.59& $-$0.67& $-$0.32& $-$0.24& $-$0.91& 58.9&1.53&S&S&S&S& 3 \\
NGC5995     &y\phantom{?}&a  & 1.78&  0.80& $-$1.23&  0.01& $-$0.65&  $<-0.75$& 19.6&6.60&S&S&S&N& 3 \\
1925$-$7245 &n\phantom{?}&a  & 1.91&  0.60& $-$0.64& $-$0.05& $-$0.62&  $<-1.44$&
56.5&2.20&S&I&S&N& 3 12 \\
IC5063      &y\phantom{?}&f  & 0.40&  1.02& $-$1.59& $-$0.28& $-$0.40&
$-$0.80&171.0&1.10&S&S&S&S&13 \\
NGC7130     &n\phantom{?}&a  & 0.75&  0.78& $-$1.06& $-$0.01& $-$0.54& $-$0.85& 95.4&2.50&S&S&S&S& 3 \\
NGC7172     &n?&a  & 0.89&  0.68& $-$0.93&  0.05& $-$0.29&  $<-0.69$&  7.6&0.07&S&S&S&N& 3 \\
NGC7479 &   ?\phantom{?}&a  & 1.11  & 0.62 & $-$0.79 & 0.06 & $-$0.14 & $<-0.17$ &7.0 & 1.00 &
S&S&S&N& 3 \\
IC5298      &n?&a  & 0.90&  0.60& $-$1.31&  0.03& $-$0.65& $-$0.53& 19.2&1.70&S&I&I&S& 3 \\
NGC7582     &n\phantom{?}&a  & 0.82&  0.35& $-$1.66& $-$0.20& $-$0.60& $-$0.94& 34.8&1.60&L&I&H&S& 3 \\
NGC7592     & &   & 0.55 & 0.40 & $-$1.18  & $-$0.20 &  $-$0.43 & &60.0 & 0.39 
&I & L & S & & 14 15 \\
NGC7674     &y\phantom{?}&eg & 0.37&  1.01& $-$0.58&  0.01& $-$0.15&
$-$1.10&584.0&1.70&S&S&S&S&1 16 \\
\end{tabular}

\caption{ Optical properties of the sample.  The second column indicates
whether we detected a HBLR in our spectropolarimetric data.  A n? indicates
that no HBLR was detected, but that we cannot absolutely rule out the
possibility that one exists with the current signal-to-noise of our data, as
discussed in Section 3.1.  A ? indicates the signal-to-noise is likely to be
insufficient to determine whether an HBLR is present or not.  A blank in this
column indicates the object was not observed.  The third column gives the
reference for the optical spectropolarimetry.  The codes are as follows: a --
this work; b -- Ruiz et al.\ (1994); c -- Antonucci \& Miller (1985); d --
Inglis et al.\ (1995); e -- Young et al.\ (1996); f -- Inglis et al.\ (1993); g
- Miller \& Goodrich (1990).  The other columns give the properties as derived
from the direct fluxed spectra.  The fourth column is the extinction, derived
assuming the intrinsic H$\alpha$/H$\beta$ ratio is 3.1.  Columns 5--9 give the
extinction corrected ratios (as $\log10$(ratio)) for (5) ${\frac{\rm 5007\AA\
[OIII]}{{\rm H}\beta}}$, (6) ${\frac{\rm 6300\AA\ [OI]}{{\rm H}\alpha}}$, (7)
${\frac{\rm 6584\AA\ [NII]}{{\rm H}\alpha}}$, (8) ${\frac{\rm 6717+6731\AA\
[SII]}{{\rm H}\alpha}}$ and (9) ${\frac{\rm 4686\AA\ HeII}{{\rm H}\beta}}$.
Column 10 gives the equivalent width of the 5007\AA\ [OIII] line in --\AA, and
column 11 the extinction corrected flux in the 5007\AA\ [OIII] line in units of
10$^{-12}$erg s$^{-1}$ cm$^{-2}$.  Columns 12--15 give the derived
classification following Veilleux \& Osterbrock (1987) from (12) ${\frac{\rm
[OIII]}{{\rm H}\beta}}$ versus ${\frac{\rm [NII]}{{\rm H}\alpha}}$; (13)
${\frac{\rm [OIII]}{{\rm H}\beta}}$ versus ${\frac{\rm [SII]}{{\rm H}\alpha}}$;
(14) ${\frac{\rm [OIII]}{{\rm H}\beta}}$ versus ${\frac{\rm [OI]}{{\rm
H}\alpha}}$; (15) presence of HeII emission.  The codes are as follows: I --
intermediate classification; L -- LINER; N -- no detection; S -- Seyfert.
Finally, the reference for the source of the spectral data is given (which in
some cases is not the same as the source of the spectropolarimetric data).
Where more than one letter is given, the first refers to the source of the line
ratios, and the second to the source of the equivalent width of the [OIII]
5007\AA\ line.  The references are as follows: (1) Osterbrock \& Martel (1993)
-- Seyfert classification is based on the probable detection of [FeX] by
Osterbrock \& Martel, which would not be expected in a typical LINER; (2)
Dahari (1985); (3) this work or other unpublished data of our own; (4) Coziol
et al.\ (1993); (5) Moran, Halpern \& Helfand (1996) classify this as a Seyfert
1.8 on the basis of weak broad H$\alpha$; (6) Kewley et al.\ 2000; (7) Ho,
Filippenko \& Sargent (1997); (8) Young et al.\ (1996); (9) De Robertis,
Hutchings \& Pitt (1988); (10) Keel et al.\ (1985); (11) Stauffer (1982); (12)
Colina, Lipari \& Macchetto (1991); (13) Colina, Sparks \& Macchetto (1991);
(14) Lonsdale, Lonsdale \& Smith (1992); (15) Kewley et al.\ (2001); (16)
Veilleux et al.\ (1995).}

\end{table}

\begin{table}
\tabcolsep=5.5pt
\begin{tabular}{lcrrrrrr}
\multicolumn{1}{c}{Name} &
\multicolumn{1}{c}{C} &
\multicolumn{1}{c}{Core Flux} &
\multicolumn{1}{c}{Total Flux} &
\multicolumn{1}{c}{Refs} &
\multicolumn{1}{c}{$F_{2-10{\rm keV}}$} &
\multicolumn{1}{c}{$N_H$} &
\multicolumn{1}{c}{Refs} 
\\
Mrk334 & 0.95 & $<5$ & 18 & a  \\
0019--7926 & &4 & 16 & ba & $<$0.10 & & 1\\
0031--2142 & &  &    &    &   0.80 & $180_{-170}^{+450}$ &2 \\
NGC1068 & 0.89 &97 & 3290 & ac$^\dagger$ & 3.50 & $>10^5$ & 3\\
NGC1143 & & 4 & 99 & bc\\
0425--0440 & && 29 & d$^*$\\
0518--2524 & 1.00 &$<3$ & 25 & ba & 4.30 & $490_{-16}^{+10}$ & 4\\
NGC4388 & 0.57 &$<6$ & 84 & ac$^\dagger$ & 12.00 &  $4200_{-1000}^{+600}$ & 5\\
IC3639 & &14 & 59 & ec$^\dagger$ & 0.13 & $>10^5$ & 6 7\\
MCG-3-34-64 & 0.67 &26 & 176 & a & 2.10 & $7600_{-1220}^{+1300}$ & 8\\
NGC5135 & 0.79&$<5$ & 133 & a & 0.20 & $>10^4$ & 9\\
NGC5194 & & & 157 & c$^\dagger$ & 1.10 & $5-10\times10^3$ & 10\\
NGC5256 & 0.16 & & 78 & c & 0.56 & $>10^5$ & 1\\
Mrk1361 && & 38 &f$^*$\\
NGC5427 & 0.04 && 20 & gh$^\dagger$ \\
1408+1347 & &$<3$ & 5 & da\\
NGC5929 & 0.08 &1 & 73 & ic$^\dagger$\\
NGC5995 & & & 21& j$^*$ & 22.00 & $86_{-30}^{+40}$ & 7\\
1925--7245 & & 33 & 180 & bk$^*$ & 0.38 & $10^3-10^4$ & 11\\
IC5063 & & 152 & 747 & dl &13.10& $2200_{-200}^{+220}$ & 9\\
NGC7130 & 0.65 &14 & 120 &ec$^\dagger$ & 0.51 & $>10^4$ & 6\\
NGC7172 & 0.43 &3 & 24&bc$^\dagger$ & 13.00 & $861_{-33}^{+79}$ & 12\\
NGC7479 & & & 76 & gm$^\dagger$\\
IC5298 & & & 25& n$^*$\\
NGC7582 & 0.62 &$<5$ & 132 &bo$^\dagger$ & 13.20 & $739_{-100}^{+146}$&9\\
NGC7592 & 0.35 && 52 & fg$^\dagger$\\
NGC7674 & 0.85 &38 & 133&ec$^\dagger$ & 0.50 & $>10^5$&13\\ 
\end{tabular}
\caption{
Data taken from the literature on the compactness of the 
10$\mu$m emission, C, compact radio flux, integrated
radio flux (both in mJy, and measured 
at 2.3GHz), observed hard x-ray flux (in units of 10$^{-12}$ergs s$^{-1}$
cm$^{-2}$ and measured in the 2--10keV band)
and inferred neutral hydrogen column density 
(in units of 10$^{20}$cm$^{-2}$)
from the x-ray data.
Where an entry is left blank no data exists in the literature for
that particular item for that galaxy.
The 10$\mu$m compactness parameter, C, is either
taken from Giuricin, Mardirossian \& Mezzeti (1995), or derived
from data presented in Maiolino et al.\ (1995), with the exception of
NGC1068 where the flux used is taken from Lumsden et al.\ (1999).
Radio data are taken from the following references: (a) Roy et al.\ (1998);
(b) Roy et al.\ (1994); (c) Rush, Malkan \& Edelson (1996);
(d) Heisler et al.\ (1998); (e) Sadler et al.\ (1995);
(f) Bicay et al.\ (1995); (g) Condon et al.\ (1990);
(h) Morganti et al.\ (1999); 
(i) Su et al.\ (1996);
(j) Condon et al.\ (1998);
(k) Roy \& Norris (1997); (l) Bransford et al.\ (1998);
(m) Condon, Anderson \& Broderick (1995);
(n) Sopp \& Alexander (1992); (o) Ulvestad \& Wilson (1984).
Where more than one reference is given, and a core flux is present,
the core flux comes from the
first listed, and the total flux from the second.  If more than one
reference is given, and there is no core flux, we have had to extrapolate
the total flux from more than one source.
Where the reference is marked with a *, the total flux has been extrapolated
to 2.3GHz assuming a spectral index of --0.75.  
The total flux is interpolated from
published data (usually at 1.5 and 5GHz) where the reference
is marked with a $^\dagger$.  The same is true for the core flux in the
case of NGC5929.  X-ray data are taken from:
(1) Risaliti et al.\ (2000); (2) Georgantopoulos (2000);
(3) Matt et al.\ (1997); (4) Kii et al.\ (1996); (5) Isasawa et al.\ (1997);
(6) Risaliti et al.\ (1999); (7) this work; (8) Ueno (1997); 
(9) Turner et al.\ (1997b);
(10) Terashima et al.\ (1998); (11) Pappa, Georgantopoulos and
Stewart (2000); (12) Guainazzi et al.\ (1998); 
(13) Malaguti et al.\ (1998).   We have made use of the catalogue
presented by Bassani et al.\ (1999) in compiling these x-ray data.
}
\end{table}

\clearpage

\begin{center}
\begin{minipage}{6.5in}{
\psfig{file=Figure1.ps,width=6.5in,angle=0,clip=}
}\end{minipage}
\vspace*{5mm}
\begin{minipage}{\textwidth}{
{\bf Figure 1:} IRAS colour-colour plots of galaxies from the survey of Kewley
et al.\ (2000).  Seyfert 1 and 2s are shown in the left-hand panels, and
starburst and LINERs (which have similar colours since LINERs tend to be
dominated by the Galactic component as far as their IRAS colours are concerned)
in the right-hand panels.  The solid line is the reddening line for Seyfert
galaxies derived by Dopita et al.\ (1998).  The dashed line is the locus of
starburst galaxies found by the same authors, and the dot-dashed line
represents the maximal extent of the region occupied by galaxies of mixed
excitation.   }\end{minipage}
\end{center}

\begin{center}
\begin{minipage}{6.5in}{
\psfig{file=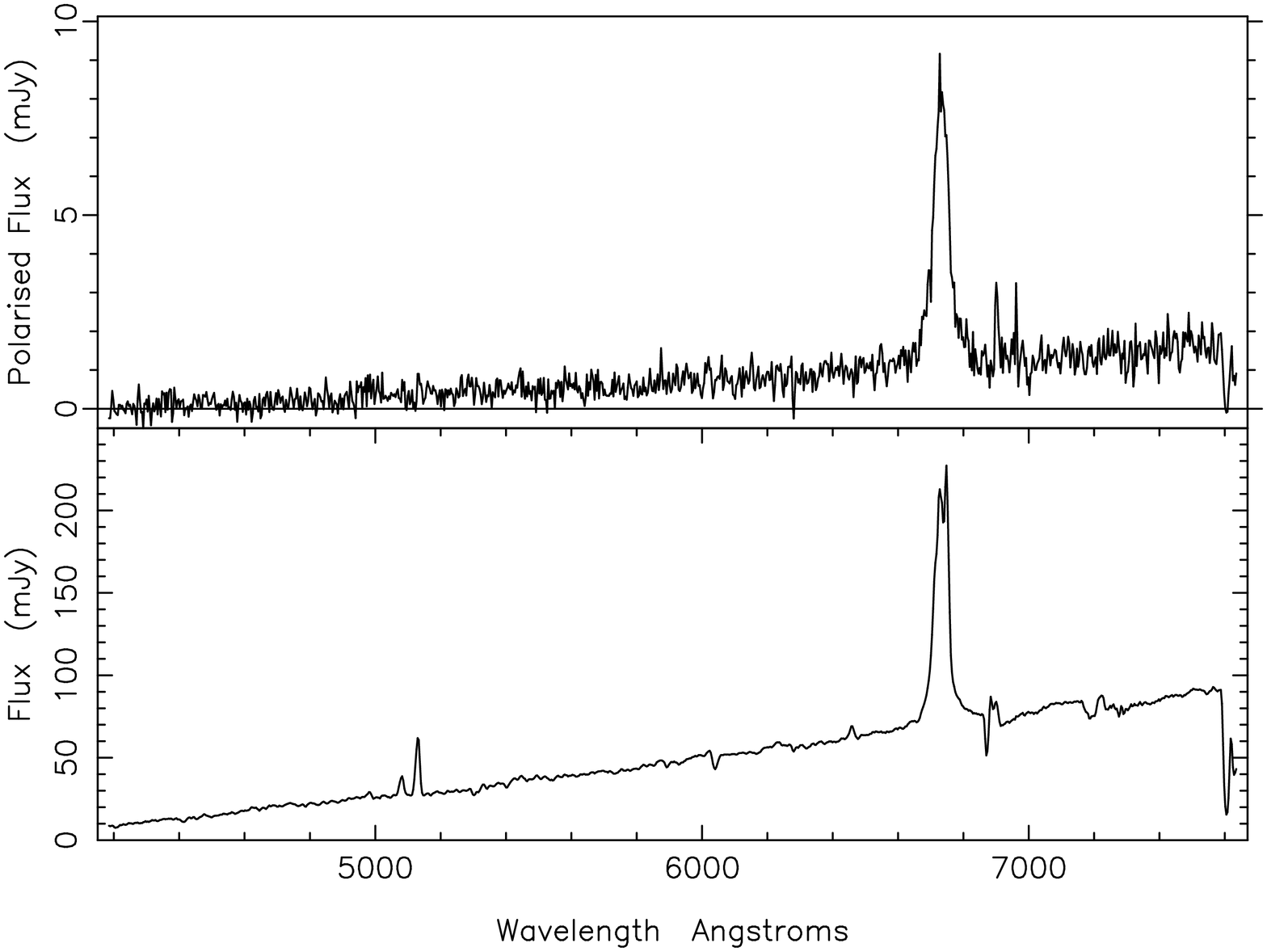,width=6.5in,angle=-90,clip=}
}\end{minipage}
\vspace*{1mm}

\begin{minipage}{\textwidth}{
{\bf Figure 2:} Spectropolarimetry of NGC5995.  The broad H$\alpha$
is clearly evident in the polarized flux.  Weak broad H$\alpha$ is
also present in the direct spectrum, leading to a classification
for this galaxy as a Seyfert 1.9.
 }\end{minipage}
\end{center}

\begin{center}
\begin{minipage}{6.5in}{
\psfig{file=Figure3.ps,width=6.5in,angle=0,clip=}
}\end{minipage}
\vspace*{5mm}

\begin{minipage}{\textwidth}{
{\bf Figure 3:} IRAS colour-colour plots as in Figure 1, but for the sample of
Seyferts considered here.  The crosses represent those galaxies in which an
HBLR has been detected, the solid circles are those galaxies without a
detection and the triangles are those galaxies for which we do not have
spectropolarimetry.  Note how the HBLRs tend to congregate on the Seyfert
reddening line, in the region of warmer IRAS colours.  Some of the Seyferts
considered clearly do not have colours dominated by the AGN from this plot,
since they have colours similar to the starburst/LINER population in Figure 1.
}\end{minipage}
\end{center}

\begin{center}
\vspace*{10mm}
\begin{minipage}{6.5in}{(a)\hspace*{5mm}\vspace*{-15mm}\hspace*{2.8in}(b)
\psfig{file=Figure4.ps,width=6.5in,angle=0,clip=}
}\end{minipage}
\vspace*{5mm}

\begin{minipage}{\textwidth}{
{\bf Figure 4:} (a) L$_{\lambda5007}$ and (b) L$_{FIR}$ as a function of IRAS mid--far
infrared colour.  There is a weak correlation present in (a) but not in (b),
suggesting that the L$_{\lambda5007}$, but not L$_{FIR}$, is a factor in the observed
colour.  The symbols are as in Figure 3.  }\end{minipage}
\end{center}

\begin{center}
\begin{minipage}{6.5in}{
\psfig{file=Figure5.ps,width=3.5in,angle=0,clip=}
}\end{minipage}
\vspace*{5mm}
\begin{minipage}{\textwidth}{
{\bf Figure 5:} The correlation between the [OIII] 5007\AA\ 
line equivalent width and the extinction corrected 
[OIII] 5007\AA\ luminosity.  The symbols are as in
Figure 3.
 }\end{minipage}
\end{center}

\begin{center}\vspace*{10mm}
\begin{minipage}{6.5in}{(a)\hspace*{5mm}\vspace*{-15mm}\hspace*{2.8in}(b)
\psfig{file=Figure6.ps,width=6.5in,angle=0,clip=}
}\end{minipage}
\vspace*{5mm}
\begin{minipage}{\textwidth}{
{\bf Figure 6:} The observed correlation between (a) the equivalent width of
[OIII] 5007\AA, and (b) the extinction corrected [OIII] 5007\AA\ to H$\beta$
ratio, and the IRAS mid--far infrared colour.  On average the HBLRs have larger
absolute equivalent widths, and higher values of the optical line ratio.  The
symbols in (a) are as in Figure 3.  }\end{minipage}
\end{center}

\begin{center}\vspace*{10mm}
\begin{minipage}{6.5in}{(a)\hspace*{5mm}\vspace*{-15mm}\hspace*{2.8in}(b)
\psfig{file=Figure7.ps,width=6.5in,angle=0,clip=}
}\end{minipage}
\vspace*{5mm}
\begin{minipage}{\textwidth}{
{\bf Figure 7:} The ratio of (a) core radio flux (essentially an AGN property),
and (b) the integrated radio flux, and the 60$\mu$m flux.  These are both
measures of how luminous the central AGN is relative to any star formation
present.  The first shows a trend for HBLRs to have higher values of the ratio.
Since the actual radio luminosities between the HBLRs and non-HBLRs are not
significantly different, this shows that the 60$\mu$m flux must be larger in
the non-HBLRs.  The trend is not as obvious in (b) especially if the radio
galaxy IC5063 is removed from consideration, showing the integrated radio
luminosity is determined in part at least by the total activity present in the
galaxy and not just the core.  The symbols are as in Figure 3.  
}\end{minipage}
\end{center}

\begin{center}
\begin{minipage}{6.5in}{(a)\hspace*{5mm}\vspace*{-15mm}\hspace*{2.8in}(b)
\psfig{file=Figure8.ps,width=6.5in,angle=0,clip=}
}\end{minipage}
\vspace*{5mm}
\begin{minipage}{\textwidth}{
{\bf Figure 8:}  The data in our sample plotted as a function of 
(a) AGN to host galaxy luminosity and obscuration and
(b) AGN luminosity and obscuration.  All of the non-HBLRs
lie at lower AGN luminosities than the HBLRs at equivalent column
densities.  This shows how luminosity and obscuration are both 
important in determining our ability to see an HBLR.
The symbols are
as in Figure 3.
 }\end{minipage}
\end{center}

\end{document}